\pdfoutput=1

\documentclass[11pt]{article}

\usepackage[]{EMNLP2023}

\usepackage{times}
\usepackage{latexsym}
 \usepackage{graphicx}
\usepackage[T1]{fontenc}

\usepackage[utf8]{inputenc}

\usepackage{microtype}

\usepackage{inconsolata}

\usepackage[shortlabels]{enumitem}
\usepackage{amsmath} 

\usepackage{xcolor}

\title{Fool Me Once? Contrasting Textual and Visual Explanations in a\\Clinical Decision-Support Setting}

\author{
Maxime Kayser$^{1*}$ \hspace{0.2cm}
Bayar Menzat$^{2}$ \hspace{0.2cm}
Cornelius Emde$^{1}$ \hspace{0.2cm} 
\\
\textbf{Bogdan Bercean}$^{3}$ \hspace{0.2cm} 
\textbf{Alex Novak}$^{4}$ \hspace{0.2cm}
\textbf{Abdala Espinosa}$^{4}$ \hspace{0.2cm}
\textbf{Bartlomiej W. Papiez}$^{1}$ \hspace{0.2cm} \\
\textbf{Susanne Gaube}$^{5}$ \hspace{0.2cm}
\textbf{Thomas Lukasiewicz}$^{1,2}$ \hspace{0.2cm}
\textbf{Oana-Maria Camburu}$^{5}$
\vspace{0.2cm}
\\
$^{1}$University of Oxford \hspace{0.2cm}
$^{2}$Vienna University of Technology \hspace{0.2cm}
$^{3}$Rayscape \hspace{0.2cm} \\
$^{4}$Oxford University Hospitals NHS Foundation Trust \hspace{0.2cm}
$^{5}$University College London
\\
$^{*}$\texttt{maxime.kayser@cs.ox.ac.uk}
}

\begin{document}
\maketitle
\begin{abstract}

The growing capabilities of AI models are leading to their wider use, including in safety-critical domains. Explainable AI (XAI) aims to make these models safer to use by making their inference process more transparent. However, current explainability methods are seldom evaluated in the way they are intended to be used: by real-world end users. To address this, we conducted a large-scale user study with 85 healthcare practitioners in the context of human-AI collaborative chest X-ray analysis. We evaluated three types of explanations: visual explanations (saliency maps), natural language explanations, and a combination of both modalities. We specifically examined how different explanation types influence users depending on whether the AI advice and explanations are factually correct. We find that text-based explanations lead to significant over-reliance, which is alleviated by combining them with saliency maps. We also observe that the quality of explanations, that is, how much factually correct information they entail, and how much this aligns with AI correctness, significantly impacts the usefulness of the different explanation types.

\end{abstract}

\section{Introduction}
    \begin{figure*}
    \centering
    \includegraphics[width=1\textwidth]{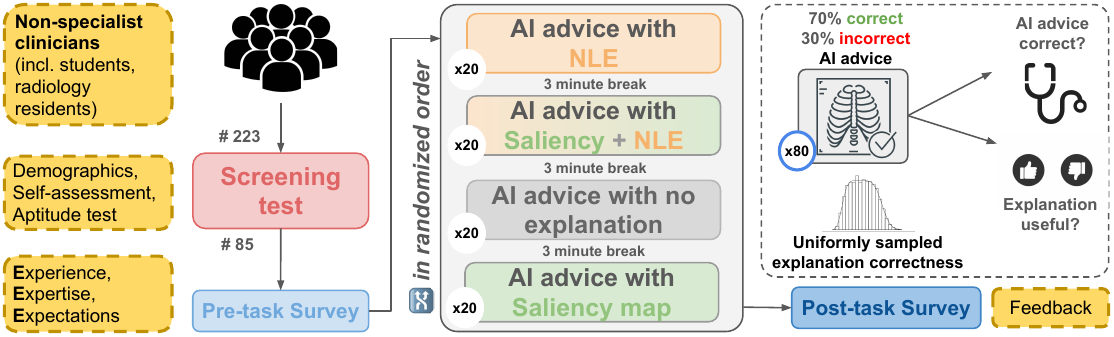} 
    \caption{The flow of the user study that every participant goes through.}
    \label{fig:design} 
\end{figure*}

AI models have progressed rapidly in recent years and are being used increasingly across various domains, including medical applications~\cite{moor2023foundation}. The communication interface of generative AI is often language-based~\cite{achiam2023gpt}, which offers a human-like mode of interaction. Some research suggests that this linguistic interface ``humanizes'' these AI systems and thereby increases reliance on them~\cite{breum2024persuasive}.

At the same time, a remaining significant barrier to the adoption and regulatory approval of deep learning models in medical imaging is the limited transparency of the reasoning processes underlying these models \cite{hassija2024interpreting}. Insufficient model robustness \cite{moss2022demystifying}, bias (algorithms are prone to amplifying inequalities that exist in the world) \cite{obermeyer_dissecting_2019, alloula2024biases}, and the high stakes in clinical applications \cite{vayena_machine_2018} are all obstacles to their wider use.

The practical utility of AI explainability methods that aim to address this remains poorly understood, as evaluating them is a challenging task.
 There can often be several correct ways to explain a decision and the criteria for judging their quality are diverse (e.g., plausibility, faithfulness, clarity \cite{jacovi_towards_2020}). Since one of the primary benefits of explanations is their utility to end-users, evaluating them with human subjects is crucial. As explanations can lead to confirmation bias and user preference frequently does not align with desired quality requirements (e.g., a user might prefer an explanation type even if they are more likely to misinterpret the AI), explanation usefulness needs to be evaluated via proxy measures~\cite{ehsan2020human, liao2022connecting, liao2021human, ehsan2021operationalizing}. Only a few studies attempt this, with some suggesting that these methods may not work as well as anticipated \cite{adebayo2018sanity, hoffmann2021looks, margeloiu2021concept, shen2020useful}.

We address this by carrying out a large-scale human subject study to evaluate the usefulness of natural language explanations (NLEs), saliency maps, and a combination of both, in the setting of imperfect AI and imperfect XAI. Saliency maps, which attribute importance weights to regions in an image, are the prevailing mode of interpretability in medical imaging \cite{van2022explainable}. NLEs, on the other hand, are becoming more widespread with recent advances in large language models~\cite{wei2022chain} and have been advocated for deployment in clinical practice~\cite{reyes_interpretability_2020}. We also study the combination of both explanation modalities, to understand if they can complement each other. We consider imperfect AI and XAI to reflect real-world applications, where both the AI predictions and the AI explanations can contain errors. Specifically, we investigate how different types of explanations, taking into account both AI and XAI \emph{correctness}, affect users in a clinical decision-support system (CDSS) environment. As we focus on AI that enhances medical practitioners~\cite{langlotz2019will, agrawal2019artificial}, rather than replaces them, our proxy for the usefulness of explanations is how much they improve human performance in human-AI collaborative chest X-ray analysis. In our study, 85 doctors and medical students analyse 80 unique images each, distributed across four different CDSS set-ups: either of the three explanation types, or the ``no explanation'' control condition. Our study design is illustrated in Fig.~\ref{fig:design}.

Our results highlight the pitfalls of language-based explanations, which lead to overreliance. Interestingly, however, saliency maps and NLEs complement each other, and their combination is the most useful explanation type. We also find that explanation correctness, and how it aligns to AI correctness, is an important factor in determining whether explanations are helpful or harmful to users. When they misalign (e.g., the AI is correct but the explanation contains a lot of incorrect information), they are detrimental to human performance, but equally, when the AI is incorrect, correct explanations mislead users into agreeing with the AI.

We find that the alignment between explanation correctness and AI correctness is critical in determining whether explanations are helpful or harmful to users. When they misalign—such as when the AI is correct but the explanation contains many inaccuracies—this negatively impacts human performance. Conversely, when they align, explanations improve our participants' task performance.
\section{Related Work}
\paragraph{XAI in medical imaging.} XAI methods can be broadly classified into post-hoc explainers and self-explaining models, i.e. approaches that either explain trained black-box AI models, or models that are inherently explainable by training and/or design. Both types have been applied widely in medical imaging applications \cite{irvin_chexpert_2019, thomas_analyzing_2019, verma_counterfactual_2020, koh_concept_2020, gale_producing_2018}. In this study we include both post-hoc explainers (saliency maps) and self-explainable models (NLEs), as well as the combination of both types.

\paragraph{Natural Language Explanations.} NLEs have been introduced as a means of providing human-understandable rationales for model predictions in computer vision~\cite{hendricks_generating_2016} and NLP \cite{camburu_e-snli_2018}. NLEs received increasing attention since then, with works aiming to benchmark and increase their \textit{plausibility} \citep{kayser_e-vil_2021, wt5}, measure their \textit{faithfulness} w.r.t.\ inner-workings of the models \citep{wiegreffe-etal-2021-measuring, atanasova-etal-2023-faithfulness, lanham2023measuringfaithfulness, siegel-etal-2024-probabilities}, and showing that they can improve model robustness \citep{he-etal-2024-using}. 
Often referred to as Chain-of-Thought (CoT) reasoning in the context of large language models (LLMs), NLEs have been used to improve reasoning capabilities~\cite{wei2022chain, zhangmultimodal}.\footnote{The concept of models generating free-text explanations before their predictions was initially introduced by \citet{camburu_e-snli_2018} and was referred to as \textit{explain-then-predict}. They also looked into how learning with NLEs can improve internal sentence representations and reasoning capabilities.} 
They have recently also been adopted in the medical domain~\cite{kayser2022explaining, chen2024chexagent}. \citet{morrison2023impact} are the first to look at the evaluation of NLEs as an interoperability tool using human subject studies. We differ by the task (safety-critical CDSS vs.\ bird classification), by looking at the combination of visual and textual explanations, and by extending explanation correctness to be continuous (rather than binary) and defined even for incorrect AI.

\paragraph{Evaluating XAI.} Evaluating AI explanations is less straightforward than evaluating, e.g., prediction performance. The lack of a unique ground truth, the wide range of interpretability goals, as well as the human-computer interaction aspect, make this more difficult. 
For these reasons, a growing body of work is evaluating XAI methods through the lens of human subject studies, generally following one of three predominant evaluation approaches described below. 

\textbf{User Preference.} Some studies directly measure human participants' preferences for XAI explanations. For instance, \citet{adebayo2020debugging} simulated a quality assurance context, requesting participants to assess the deployment readiness of AI algorithms, which came with different kinds of explanations.
However, \citet{hase2020leakage} demonstrated that user preference does not correlate with how well users can predict model behavior, a proxy for how transparent the model is.
Additionally, there are concerns that humans are prone to confirmation bias, i.e., focusing on evidence that confirms preexisting expectations in a model explanation \cite{wang2019designing}. There is also evidence that XAI methods can unreasonably increase the confidence in a model's prediction \cite{kunkel2019let, schaffer2019can, ghassemi2018clinicalvis, eiband2019impact}.

\textbf{Model Predictability.} Arguably, the closest proxy for \emph{full} model transparency is to measure how well humans can predict a model's predictions on unseen data. If users can correctly predict the model on all unseen data, it means the model is entirely transparent to them. 
While some works opt for this method on simplified problems~\cite{alqaraawi2020evaluating, colin2022cannot, yang2019evaluating, shen2020useful}, its applicability to radiology is limited, as predictions are highly nuanced and explanations are complex and label-specific.

\textbf{Human-AI Team Performance.} Another approach to evaluate the usefulness of XAI explanations is to measure how much they improve human performance in the AI-human collaborative setting. The goal of XAI in this setting is to guide the user to appropriate evidence when the model is correct, or shed light on faulty AI reasoning when it is wrong. \citet{chu2020visual} measured the impact of XAI methods in helping users predict age given images of human faces. \citet{kim2022hive} analyzed performance changes in a bird classification task under the guidance of various XAI techniques.
In clinical applications, where practitioners see a need for explanations to justify ``their decision-making in the context of a model’s prediction'' \cite{tonekaboni2019clinicians}, this evaluation method is particularly well suited and hence also used in this work.

\paragraph{Evaluating XAI in CDSSs.} 

In medical imaging, where concerns around safety and trust make autonomous deployment of AI models challenging, there is an emphasis on how AI can collaboratively support medical professionals. CDSSs, where AI models offer recommendations to humans for specific tasks, are a common form of human-AI collaboration in clinical practice.

Existing studies investigate this form of human-AI interaction by looking at how the sequential order of human and AI decisions affect performance \cite{fogliato2022goes}, what influence the assertiveness of AI suggestions has \cite{calisto2023assertiveness}, or which kind of users benefit the most from it \cite{gaube2023non}. A recent large-scale study conducted by \citet{NBERw31422} shows that, in most cases, human performance is enhanced when using CDSSs.

Few works have looked at the usefulness of XAI in clinical applications. \citet{du2022role} consider a simple, 5-feature set-up to compare explanation-based and feature attribution methods in a CDSS setting. \citet{rajpurkar2020chexaid, ahn2022association} provide visual explanations when evaluating the usefulness of a CDSS, but they do not look at the effect that XAI explanations had. \citet{gaube2023non} find that visual explanations improve the diagnosis performance for non-task experts, but they do not compare it to other XAI methods. \citet{tang2023impact} look at AI tools for lung nodule detection in chest X-rays and find that localization maps do not improve performance.
In contrast to previous work, we are the first to consider language-based explanations, compare the effect of different explanation types, and take into account their interaction with diagnosis and explanation correctness in a clinical context.

\begin{figure*}[h] 
    \centering
    \includegraphics[width=1\textwidth]{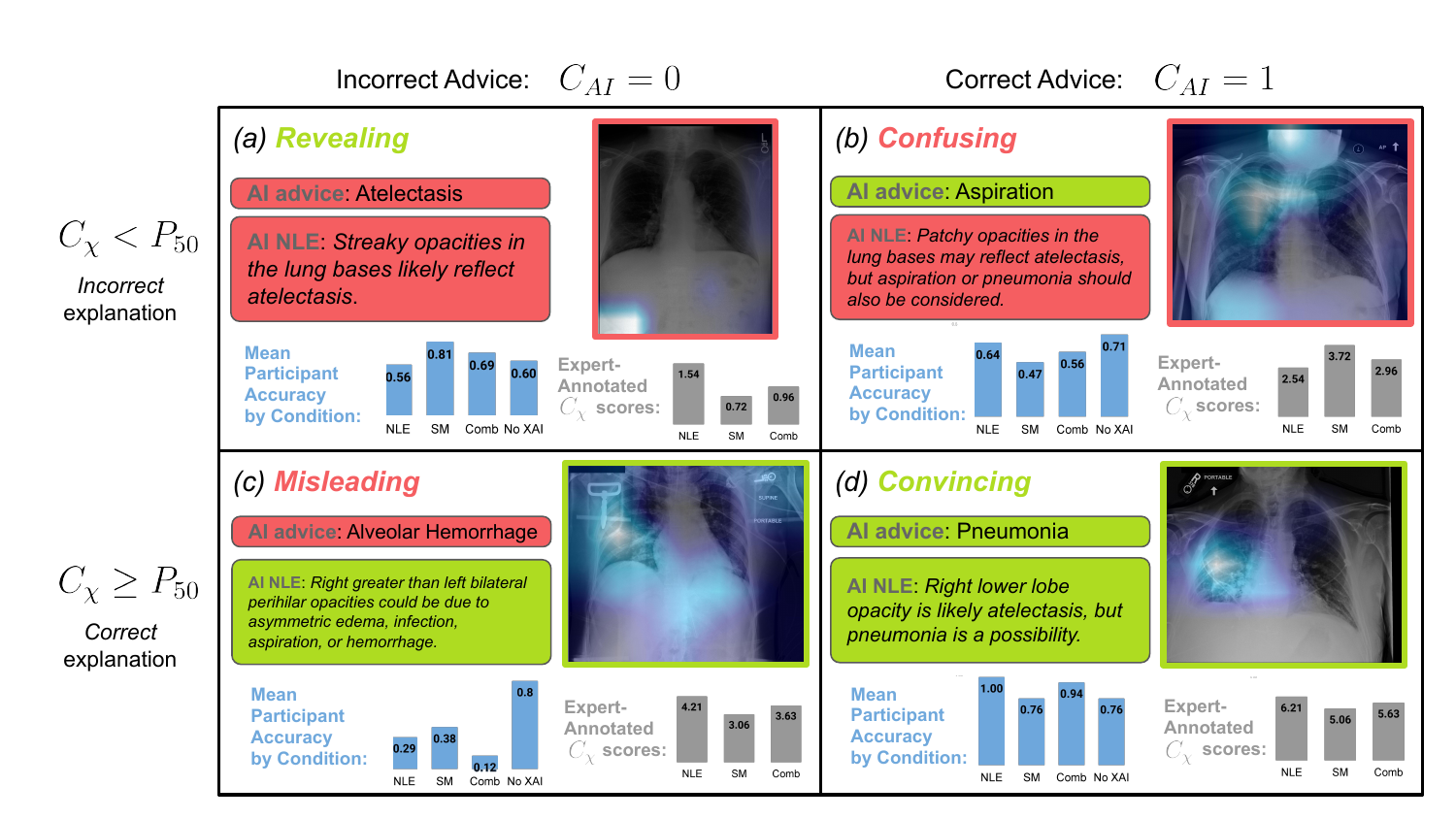}
    \caption{(a) \emph{Revealing} ($C_{AI} = 0$, low $C_\chi$): The AI incorrectly suggests atelectasis, but the poorly rated explanations help clinicians identify the error, leading to higher accuracy compared to relying on the AI prediction alone. (b) \emph{Confusing} ($C_{AI} = 1$, low $C_\chi$): The AI correctly identifies aspiration but provides low $C_\chi$ explanations3, leading to lower diagnostic accuracy compared to the No XAI setting. (c) \emph{Misleading} ($C_{AI} = 0$, high $C_\chi$): The AI incorrectly suggests alveolar haemorrhage but provides highly rated explanations, misleading participants to agree with the incorrect AI when explanations are provided. (d) \emph{Convincing} ($C_{AI} = 1$, high $C_\chi$): The AI correctly identifies pneumonia and provides highly rated explanations, resulting in high diagnostic accuracy, especially for NLEs.}
    \label{fig:four_quadrants}
    \end{figure*}
\section{Methods}
    We designed a study to evaluate the usefulness of NLEs, saliency maps, and their combination in a clinical decision-support context. We also control for AI advice correctness $C_{AI} \in \{0,1\}$ and explanation correctness $C_\chi \in [1,7]$. Our main research question is how different explanation types, in the context of advice correctness $C_{AI}$ and explanation correctness $C_\chi$, affect human performance on the task of classifying chest X-rays, where explanation \emph{usefulness} equates by human performance.

\textbf{Definition of explanation correctness $C_\chi$:} $C_\chi$ captures to what extent the information provided in an explanation is clinically, factually correct. An explanation can be incorrect (i.e., contain a lot of incorrect information) even when the AI prediction was correct, and vice versa, similar to definitions from~\citet{honovich2022true} and~\citet{zhang2020optimizing}.
Note that this is different from other explanation criteria such as \textit{faithfulness} (i.e., how ``accurately it reflects the true reasoning process of the model'') and \textit{plausibility} (i.e., how convincing the explanation is to humans) \cite{jacovi_towards_2020}.

We obtain the ground-truth for both advice and explanation correctness from annotations by three expert radiologists. For each of the three explanation scenarios, $C_\chi$ is rated on a 7-point Likert scale. The evaluation interface given to the annotators is shown in Figure~\ref{fig:annotators} in the Appendix.

\subsection{Study Overview}

Our pre-registered, IRB-approved\footnote{osf.io/nf52s; Approval Nr. CS\_C1A\_23\_018\_001} user study involves 85 clinical participants and was developed through iterative pilot studies and consultations with expert clinicians. We use a human-AI collaborative setup to evaluate the \emph{usefulness} of explanations in terms of their ability to help a user discern whether a model's prediction is correct or not. We present both quantitative and qualitative measurements. The study design is outlined in Fig.~\ref{fig:design}. 

Our CDSS provides a suggestion (the \emph{AI advice}) for each image, consisting of a single radiographic finding predicted by the AI. To simplify our design, we focus on one finding per image, and communicate to participants that this is neither necessarily the only nor most important finding. We simulated an environment where the model has an accuracy of 70\%, to strike a balance between having a reasonable representation of correct and incorrect model predictions and not making the model appear overly unreliable. We also sample image-explanation pairs to ensure that the overall distribution of $C_\chi$ scores is as uniform as possible (so that all $C_\chi$ levels are well represented), see Figure~\ref{fig:ec_scores} in the Appendix.

We study the following four conditions: (i) $\chi_{\text{None}}$ (participants receive the AI model’s advice without any explanation), (ii) $\chi_{\text{SM}}$ (participants receive the model’s advice and a saliency map), (iii) $\chi_{\text{NLE}}$ (participants receive the model’s advice and an NLE), (iv) $\chi_{\text{Comb}}$ (participants receive the model’s advice, a saliency map, and an NLE). A screenshot of the user interface is shown in Figure~\ref{fig:screen2}. For each condition, participants are shown 20 cases, which consist of a chest X-ray, the patient context, the AI advice, e.g., ``Pneumonia'', and a condition-specific explanation. They are then asked to express their agreement with the AI advice (``Not present'', ``Maybe present'', or ``Definitely present''). We also ask them whether they found the explanation useful in their decision-making (e.g. ``How useful was the AI model's explanation in helping you decide whether the AI was right or wrong in suggesting pneumonia.''). This is meant to encourage them to engage with the explanation and it enables us to quantify the relationship between \emph{perceived} and \emph{actual} explanation usefulness.

To mitigate order effects and user fatigue, we randomize the order of the conditions for each participant. We also enforce three-minute breaks between each condition, where we give participants the option to follow a guided meditation. We also emphasize multiple times that the users are engaging with different AI models in each condition, to avoid carry-over effects where a person's engagement with explanation type A affects their perception of the CDSS and therefore their subsequent engagement with explanation type B. Finally, we introduce an incentive of doubling the compensation for participants who perform in the top 20\%. This is to ensure that users are dedicated throughout the 80 cases. At the end of the four conditions, users fill out a post-study survey. Here we ask them about their experience with the different AI explanations and measure how their attitude towards AI has been affected. The entire task is conducted online via a custom streamlit platform that we make publicly available for future use.\footnote{\url{https://github.com/maximek3/fool-me-once}}

\subsection{Participant Recruitment}

As we aim to study the effect of different explanation types in an imperfect (X)AI setting, we recruit participants with foundational competence in reading chest X-rays, who are knowledgeable enough to not rely wholly on the AI system, but are still likely to engage with the AI's predictions and explanations. 
Furthermore, CDSSs are generally seen as most useful for people who have medical training but are not experts in the task at hand \cite{bussone2015role}. This is particularly relevant in scenarios where there is a scarcity of expert radiologists~\cite{mollura2020artificial}, and non-expert clinicians benefit from collaborating with AI systems \cite{gaube2023non}. For these reasons, our primary target group for this study are medical students and doctors who have undergone training in reading chest X-rays, but who are not specialist radiologists. Our sample size was estimated via a power simulation based on several pilot studies. More information is provided in Appendix~\ref{app:participants}.

\subsection{Model Implementation}

In the eyes of our participants, they are presented with four different AI models throughout the study. In reality, to ensure comparability, the backbone vision classifier is the same for all images. We train a transformer-based vision-language model (VLM) following the Ratchet architecture as in~\citet{kayser2022explaining}. It consists of a DenseNet vision encoder \cite{huang2017densely} that generates 7x7 1024-dimensional feature maps of the image. These are then both pooled to perform multi-label image classification and flattened to be given as prefixes to a transformer decoder for NLE generation. The NLE is further conditioned on the predicted label, i.e. the VLM predicts the class and generates an NLE conditioned on the prediction and the learned image representation. 

From this VLM we then extract the four models introduced in our four conditions.  For $\chi_{\text{None}}$ we only use the backbone Densenet, $\chi_{\text{SM}}$ consists of the backbone Densenet and saliency maps extracted from this backbone, $\chi_{\text{NLE}}$ uses the entire VLM, without saliency maps, and $\chi_{\text{Comb}}$ adds saliency maps to the VLM.

The VLM was trained on the MIMIC-NLE dataset \cite{kayser2022explaining}, containing both findings (i.e., diagnoses) and NLEs. 
The NLEs are all directly extracted from radiology reports that were recorded during routine clinical practice. Each NLE links a finding to its evidence in a radiographic scan, including details about location, size, severity, certainty, and differential diagnoses. Examples of model-generated NLEs are shown in Figure~\ref{fig:four_quadrants}. The model obtained a weighted AUC of 0.75. Note that the main purpose was not to maximize model performance. Instead, we specifically focus on the case of imperfect AI, where a model, for various reasons, such as limited or biased data, does not perform optimally. 
Nonetheless, our model still performs favorably on existing benchmarks, ensuring that our model and the generated explanations are of a realistic standard \cite{irvin_chexpert_2019}.

The model learns to generate NLEs in a supervised way. Therefore, the generated NLEs capture the nuances around assertiveness and the certainty of findings that naturally occur in clinical practice. For this reason, we consider assertiveness an integral part of the NLEs, as opposed to a design factor that can be studied by itself \cite{calisto2023assertiveness}.

We implement Grad-Cam \cite{selvaraju_grad-cam_2017} following \citet{jacobgilpytorchcam} to obtain saliency maps. We chose Grad-CAM as it is widely used and previous work has shown that out of the commonly used saliency techniques, it is the most accurate one for medical imaging \cite{saporta2022benchmarking}. We have also qualitatively verified it by comparing it to Grad-Cam++, HiResCam, AblationCAM, and XGradCAM~\cite{jacobgilpytorchcam}. 

\subsection{Obtaining the Study Samples}

Even though our chest X-rays are paired with human-written radiology reports, we follow existing work \cite{gaube2023non, ahn2022association, seah2021effect} and have three experienced radiologists annotate the correctness of our AI advice and explanations. Details on this process are in Appendix~\ref{app:annos}. 

We annotated 160 examples, from which we carefully selected 80 cases to control the share of incorrect predictions by each class, ambiguity, and the distribution of $C_\chi$ scores. We include the radiographic findings pneumonia, atelectasis, pulmonary edema, fluid overload/heart failure, aspiration, and alveolar haemorrhage. More information on the case selection process is provided in Appendix~\ref{app:samples}.
\section{Results}
    \subsection{Statistical Model}

We model our results using a Generalized Linear Mixed-Effects Model (GLMM) that predicts human accuracy for each instance. We chose GLMMs because they offer a flexible and robust framework to handle non-normally distributed outcome variables and account for both fixed and random effects. The below model carefully accounts for our complex study design, including the triple interaction terms (explanation type, correctness, and advice correctness) and missing values by design (no explanation correctness for $\chi_{\text{None}}$). We follow best practices from~\citet{koch2023tutorial}. We define explanation type as $\chi$. The GLMM is given below:

\begin{equation}\label{eq:model-definition}
\begin{aligned}
l_{ij} = & \beta_0 \\
& + \beta_a C_{AI}  \\
& + \beta_t \chi \\
& + \beta_{t \times a} (\chi \times C_{AI}) \\
& + \beta_{t \times e} (\chi \times C_\chi) \\
& + \beta_{t \times e \times a} (\chi \times C_\chi \times C_{AI}) \\
& + u_{Participant} \\
& + u_{Image}
\end{aligned}
\end{equation}

This model predicts the log-odds of the human accuracy $l_{ij}$ for the $i$-th participant on the $j$-th image. As fixed effects, we consider advice correctness $C_{AI} \in \{0,1\}$, explanation type $\chi \in \{\chi_{\text{None}}, \chi_{\text{SM}}, \chi_{\text{NLE}}, \chi_{\text{Comb}}\}$, explanation correctness $C_\chi \in [-3,3]$ (mean-centered from 7-point Likert scale), and different interactions of these effects. As random effects, we include the participants $u_{Participant}$ (who can have different skill levels) and the images $u_{Image}$ (which can have different difficulty levels). 

\begin{table}[]
    \centering
    \caption{Our framework for classifying AI explanations. Green squares are \textcolor[HTML]{AEDC21}{\emph{insightful} explanations}. Red squares are \textcolor[HTML]{F55E61}{\emph{deceptive} explanations}. $P_{50}$ denotes the 50-th percentile. Illustrative examples for each quadrant are shown in Figure~\ref{fig:four_quadrants}.}
    \begin{tabular}{lcc}
    \hline
             & $C_{AI} = 0$ & $C_{AI} = 1$ \\
    \hline
    $C_\chi < P_{50}$ & \textcolor[HTML]{AEDC21}{Revealing} & \textcolor[HTML]{F55E61}{Confusing} \\
    $C_\chi \geq P_{50}$ & \textcolor[HTML]{F55E61}{Misleading} & \textcolor[HTML]{AEDC21}{Convincing} \\
    \hline
    \end{tabular}
\label{tab:quadrants}
\end{table}

Rationales for the different interaction terms is given below:
\begin{itemize}
    \item $\chi \times C_{AI}$: We assume that different explanation types have a different impact on human accuracy when advice is correct or incorrect. For example, explanation types prone to confirmation bias will have a particular effect when the advice is incorrect.
    \item $\chi \times C_\chi$: Note that we do not include $C_\chi$ as a main effect. This is because $C_\chi$ between different explanation types are not directly comparable (e.g. NLEs contain more specific information and therefore can contain both more correct information and more false information). Therefore we consider $C_\chi$ as a type-specific metric and need to include the interaction term.
    \item $\chi \times C_\chi \times C_{AI}$: We need to model this interaction as $C_\chi$ strongly correlates to $C_{AI}$. This is because incorrect advice generally has explanations with a lot less correct information, and therefore $C_\chi$ is much lower when $C_{AI}=0$.
\end{itemize}

We fit this model on our data to interpret the effects and test different hypotheses that align with our research question. Due to data dependencies, we opted for a mixed model approach to test adjusted means, rather than performing inferential statistics on observed means. Model parameters of this three-way full factorial model GLMM are hard to interpret because the probability $l_{ij}$ is the log-odds of human accuracy, the random effects, and because our various interaction terms that making hard to isolate different factors. For this reason, we do not directly discuss effect sizes and significance values for individual model terms. For example, $\beta_{\chi_{NLE}}$ only represents the log-odds of the human accuracy when $C_{AI} = 0$ and $C_\chi = 0$, not the effect of $\chi_{NLE}$ as a whole. Instead, we focus on using our model to predict human accuracies on our observations and test differences via contrasts. The majority of results in our paper, such as in Fig.~\ref{fig:main_results}, are based on hypothesis testing using the \textit{marginaleffects} package~\cite{marginaleffectsForthcoming}.

We test the model statistically and find that both random and fixed effects should be included. In particular, we perform a likelihood ratio test (LRT) between the model in Eq. \eqref{eq:model-definition} and a baseline model disregarding explanation correctness and interactions. We find that the full model yields a significantly better fit $\chi^2_{12}=28.21$, $p=.005$ (we provide more details in the Appendix \ref{app:model-selection}).

\begin{figure}[ht]
\centering
\includegraphics[width=0.45\textwidth]{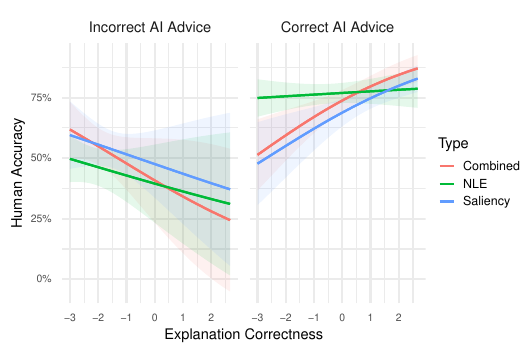}
\caption{Human accuracy given $C_{AI}$ and $C_\chi$, predicted with the model \eqref{eq:model-definition}.}
\label{fig:ec_lines}
\end{figure}

Our GLMM in Eq.~\ref{eq:model-definition} was also used for our power analysis to estimate the sample size. We estimated effect sizes via multiple pilot studies and related work~\cite{gaube2021ai}. Following this rigid procedure ensured that our study was well-powered and that our model assumptions were validated prior to data collection.

\subsection{Post-Survey Insights}

Before delving into the statistical findings we first look at the outcome of our post-task survey, where we asked users about their experience with the different explanation types. There is a strong trend of NLEs being preferred the most, and saliency maps the least, as shown in Table~\ref{tab:ranking}. Participants perceived the model with saliency maps to be on average 17\% less accurate than the model with NLEs, even though all models had the same accuracy by design.
Each explanation type was also evaluated across five key characteristics of explanations, with language-based explanations scoring the highest on all five, as shown in Fig.~\ref{fig:spider} (the questions can be found in Appendix~\ref{app:survey}). NLEs are preferred across all characteristics. In the remainder of this paper, we will look at whether this preference aligns with \emph{usefulness}.

\begin{table}[]
    \centering
    \caption{Preference ranking of models.}
    \small  
    \begin{tabular}{lccccc}
    \hline
             & $\mu$Rank & \#1 & \#2 & \#3 & \#4 \\
    \hline
    NLE      & 1.85     & 38.9\%     & 38.9\%     & 20.0\%     & 2.21\%      \\
    Comb. & 2.05     & 40.0\%     & 23.3\%     & 27.8\%     & 8.90\%      \\
    No XAI   & 2.98     & 14.4\%     & 21.1\%     & 16.7\%     & 47.8\%     \\
    SM & 3.11     & 6.72\%      & 16.7\%     & 35.6\%     & 41.1\%     \\
    \hline
    \end{tabular}
\label{tab:ranking}
\end{table}

\begin{figure}[ht!]
\centering
\includegraphics[width=0.45\textwidth]{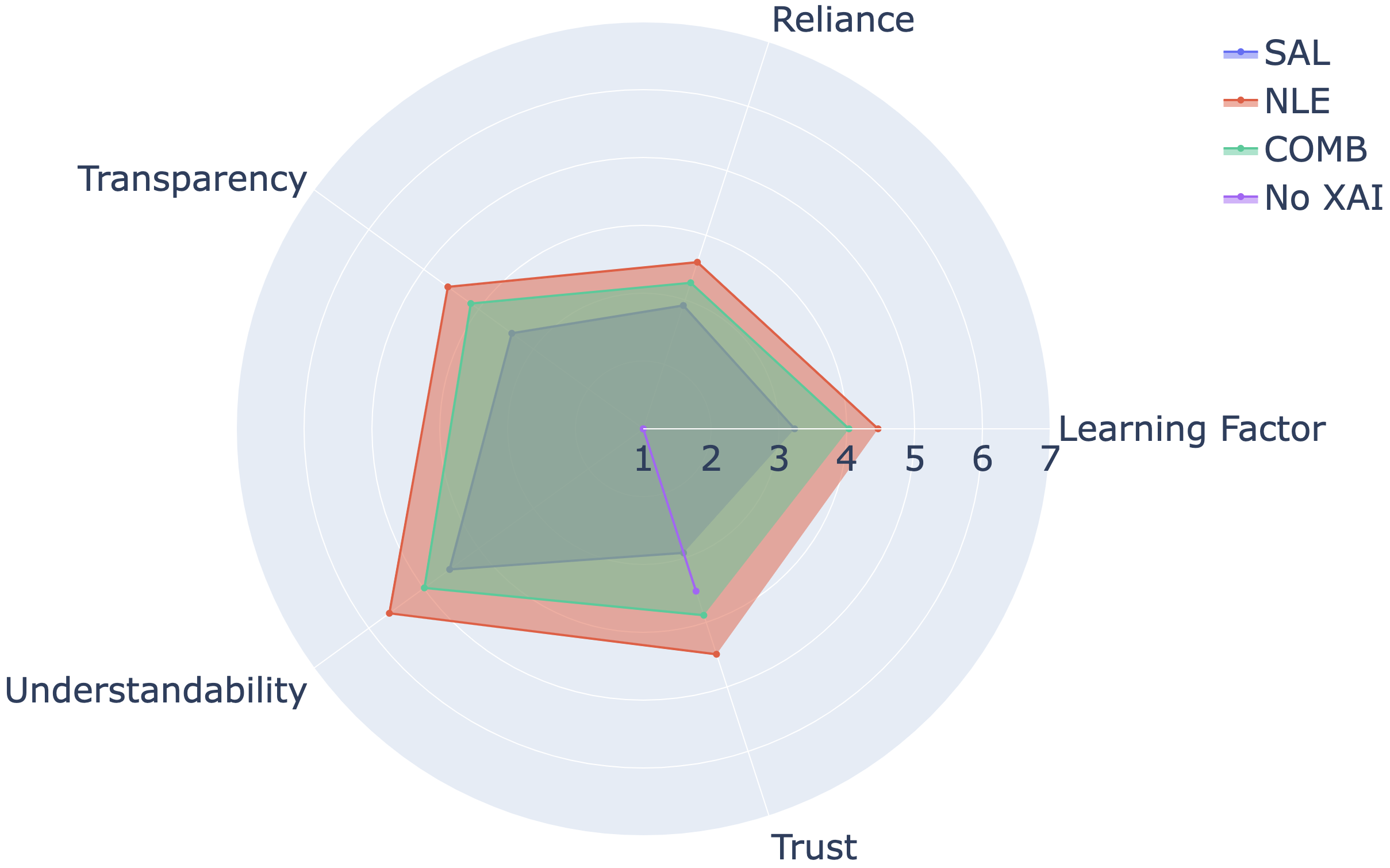}
\caption{Five attributes of explainability methods, ranked on a 7-point Likert scale.}
\label{fig:spider}
\end{figure}

\subsection{Main Results}\label{RQs}

\begin{figure*}[h]
\centering
\includegraphics[width=1\textwidth]{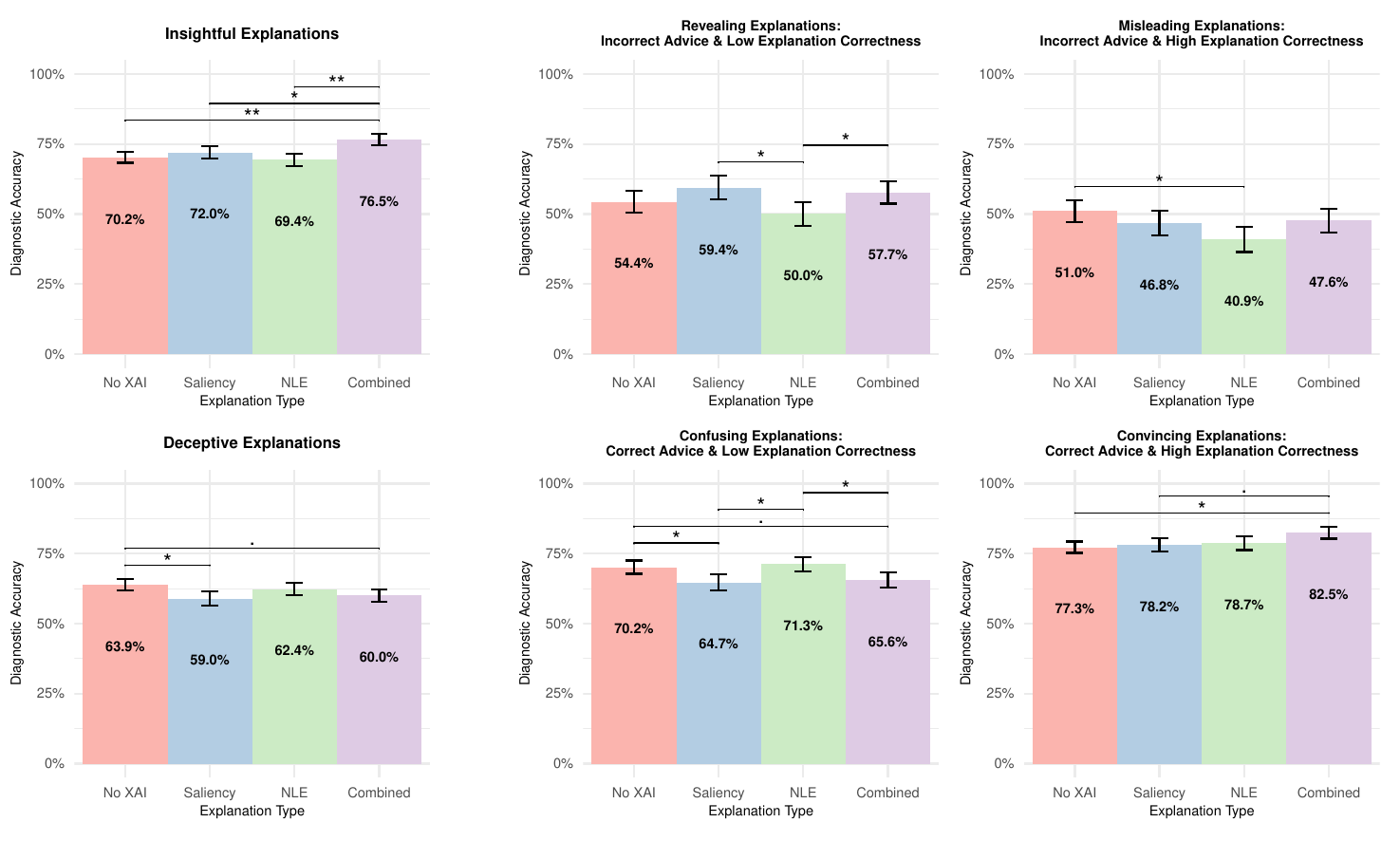} 
\caption{The bar charts represent model-based predictions of human accuracy under different conditions. For example, the model predicts a 76.5\% ``expected probability'' of correct user decisions for ``insightful explanations'' with NLEs (top-left plot). $p$-values are derived from hypothesis testing, comparing human accuracy between explanation types for specific data subsets. The error bars represent standard errors. $\cdot$, *, ** ($p<0.1$, $0.05$, $0.01$)}
\label{fig:main_results} 
\end{figure*}

To capture the various ways in which advice and explanation correctness can interact, we propose the framework described in Table~\ref{tab:quadrants} to interpret advice and explanation correctness. Example cases for the different interaction types are shown in Fig.~\ref{fig:four_quadrants}. We split explanations into incorrect ($C_\chi$ in lower 50\% percentile) and correct ($C_\chi$ in upper 50\% percentile). The results are in Fig.~\ref{fig:ec_lines} and~\ref{fig:main_results}.

\paragraph{NLEs on their own lead to overreliance.}

Across all $C_{AI}$ and $C_\chi$ scores, differences between our four conditions cancel each other out and we observe no significant differences (see Figure~\ref{fig:hyp123} in the Appendix). However, for incorrect advice, there is a significant drop in human accuracy for NLEs compared to combined ($-7.3$\%, $p<.05$) and saliency maps ($-6.2$\%, $p<.05$). This suggests that NLEs make people more likely to agree with the AI when it is actually incorrect. While alarming, this is not unsurprising given that participants rated the NLE model much higher in the post-study survey, suggesting that they overestimate that model and hence overrely on it. Especially when $C_\chi$ is comparatively high but the AI advice is incorrect, people are 10.1\% ($p<0.05$) more likely to agree with the AI than without explanation. This also means that for the scenario of correct advice and comparatively low $C_\chi$ explanations, NLEs lead to higher performance (6.6\%, $p<.05$ vs. saliency maps and 5.7\% $p<.05$ vs. combined), as people are more likely to agree with low $C_\chi$ NLEs than other explanation types with low correctness. Overall, people agree with the AI 67.3\% of the time when it is accompanied by an NLE, compared to 63.8\% on average for the other explanation types. This could suggest that the assertiveness~\cite{calisto2023assertiveness} and/or human-like \cite{breum2024persuasive} nature of language-based explanations could lead people to overly trust and rely on AI. 
    
\paragraph{$C_\chi$ needs to \emph{align} with $C_{AI}$:} Our results show that insightful explanations, i.e., where $C_\chi$ aligns with $C_{AI}$, are helpful in a decision-support setting. 
Figure~\ref{fig:ec_lines} illustrates how higher $C_\chi$ scores harm human accuracy when the AI prediction is incorrect (\emph{deceptive} explanations) and benefits human accuracy when the AI advice is correct (\emph{insightful} explanations). These effects are less strong for NLEs than for the visual methods. 

In Figure~\ref{fig:main_results}, we look at human accuracy by explanation type for the four $C_\chi$ scenarios described earlier. To obtain human accuracy for "No XAI", where we do not have explanations correctness scores, we simply consider all the images where the average of all other explanation correctness scores is in the upper half or lower half.

We observe that, as a general trend, human accuracy is harmed when explanations are \emph{deceptive}, and people would be better off seeing no explanation. For saliency maps, human accuracy goes down 4.9\% ($p<.05$) when $C_{AI}$ and $C_\chi$ do not align. For combined explanations, it goes down 3.9\% ($p=.06$). On the contrary, for insightful explanations, human accuracy goes up 4.3\% ($p<.005$) for combined explanations. These effects are not seen for NLEs, suggesting that the visual explanations are more helpful to users to discern whether an AI's decision-making is flawed.

\paragraph{When aligned, combine saliency maps and NLEs.}

For insightful explanations, where correctness aligns with AI correctness, combining saliency maps and NLEs provides significant improvements compared to the other conditions: 6.3\% ($p<.005$) over ``No XAI'', 7.1\% ($p<.005$) over NLEs alone, and 4.5\% ($p<.05$) over saliency maps alone. This suggests that participants are able to integrate the information from both visual and textual cues to identify when an AI is wrong or right. Interestingly, even though insightful NLEs on their own are worse than ``No XAI'', combining them with visual explanations leads to a significant boost. 

We ensure the robustness of our results by pre-registering our study, aligning with best practices to avoid p-hacking~\cite{wicherts2016degrees}, having a rigorous model selection process, guided by AIC and BIC in addition to likelihood ratio tests, and by minimizing the number of subsequent tests. Finally, we report effect sizes and confidence intervals over p-values where possible, focusing on practical significance~\cite{nakagawa2004farewell}. Multiple-testing adjusted results in Figure~\ref{fig:main_results_adj} in the Appendix.

\subsection{Exploratory Results}

In addition to our main research question, we also measured ``perceived usefulness'', ``decision speed'', and ``positive certainty''. We summarize the most important findings here and provide more details and analysis in Appendix~\ref{app:exploratory-analysis}.

\paragraph{Perceived usefulness.} Perceived usefulness is a subjective measure of how useful participants find an explanation (7-point Likert scale response to ``How useful was the AI model’s explanation in helping you decide whether the AI was right or wrong in suggesting (e.g.) pneumonia''). This allows us to measure subjective preference on a per-instance level and juxtapose it to ``objective'', actual usefulness. We find that NLEs, in line with our post-study survey, are consistently rated the most useful (Fig.~\ref{fig:pu_main}). Even though this contrasts actual usefulness (human accuracy), there is no significant difference in how perceived and actual usefulness misalign between explanation types. Our assumption that low $C_\chi$ saliency maps help users detect when the AI is wrong is confirmed in Fig.~\ref{fig:pu_lines}.

\paragraph{Positive certainty.} We define positive certainty as the share of times participants say a finding is ``Definitely'' instead of ``Maybe present'' (for negative, we only have ``Not present'', so we cannot measure the degree of certainty). We find that it is hard to predict positive certainty and that it does not vary significantly by explanation type. Unsurprisingly, it is highest for \emph{convincing} explanations (Fig.~\ref{fig:pc_charts}).

\paragraph{Decision speed.} Decision speed is the time taken to provide an answer for a single chest X-ray. Decision speed increases significantly with the level of complexity of the explanations, going from 36.0 seconds for no explanation to 39.6 for saliency maps, 42.8 for NLEs, and 43.1 for combined (Figure~\ref{fig:ds_charts}). Explanation correctness and the quadrants have no significant effect on decision speed.

\section{Summary and Outlook}
In this work, we conducted a large-scale user study simulating a real clinical decision support set-up and included in-domain, clinical experts. We juxtaposed textual (NLEs) and visual (saliency maps) explanations, and found that NLEs lead to severe overreliance, but can be helpful when combined with visual explanations. We also show that alignment between explanation and advice correctness is a strong predictor for explanation usefulness. This study sheds light on the pitfalls of convincing-sounding language-based explanations and we hope it enables future research on optimizing such explanations to lead to safe use of AI.

\section*{Limitations}
Our study provides a snapshot of how users engage with AI and its explanations in our experimental set-up. Even though we tried our best to replicate real clinical practice, including with the use of incentives, our study cannot fully replicate the conditions under which clinicians work. This is also not a longitudinal study, meaning we do not explore how interaction with models and explanations change over time. It is worth noting that recruitment biases such as self-selection can impact the participants who chose to engage in this study. Even though our cohort of participants is fairly diverse, it is still most likely not representative of the global population as a whole.

\section*{Acknowledgments}
We want to sincerely thank Guy Parsons, Lize Alberts, and Florian Pargent for their helpful discussions. Maxime Kayser is part of the Health Data Science CDT at the University of Oxford. Oana-Maria Camburu was supported by a Leverhulme Early Career Fellowship. Thomas Lukasiewicz and Maxime Kayser 
were also supported by the AXA Research Fund.
We would also like to thank all the participants in our study, amongst others: Catalina Beatrice Cojocariu, Fatema Aftab, Veronica-Maria Urdareanu, Dr. Vani Muthusami, Necula Anca Mihaela, Valentin-Razvan Avram, Dr. Chloe Panter, Montague Mackie, Dr. Malacu Oana-Alexandra, Varga Alexandra, Catarina Santos, Iulia Ilisie, Kevin A. Militaru, Nucu Iuliana Alexandra, Mirela Moldovan, Anam Choudhry, Dr. Alexandrescu Ionela-Roxana, Ana Hârlău, Dr R. W. Mifsud, Fisca Sorina Madalina, SimileOluwa Onabanjo, Adnan Anwar, Lucia Indrei MD, Păcuraru Daniela-Sena, Bilal Qureshi, Oana Andreea David, Jamie Brannigan MA MB BChir, Michael Watson, Popa Cosmin-Gabriel, Iulia-Gabriela Ghinea, Michael Milad, Sanskriti Swarup, Faisal Shaikh, Mouna Mayouf, Kejia Wu, Steren Mottart, Katerina Gramm, RTS Alkaissy, Dr. Da Cloete, Diana-Andreea Ilinca, Humayun Kabir Suman, Robyn Gould, Jade Williams, Sofia Baldelli, Stefana Grozavu, Isaac K. A. Nsiah, Stefania-Irina Hardulea, Aleksander Stawiarski, Chidinma Udojike, Tom Syer, Nicoleta Ioana Lupu, Dr. Edmond-Nicolae Bărcan, Botez A.M., Baboi Delia Andreea, Isabelle Zou, Mleziva Bianca, Charles Hillman, Dr. Iain Edgar, Dr. Olanrewaju Abdulrazaq, Kriti Sarin Lall, and Dr. Fanut Luciana. The remaining participants remained anonymous.

\bibliography{custom}
\bibliographystyle{acl_natbib}

\clearpage
\newpage
\appendix
    \begin{figure*}[ht]
\centering
\includegraphics[width=1\textwidth]{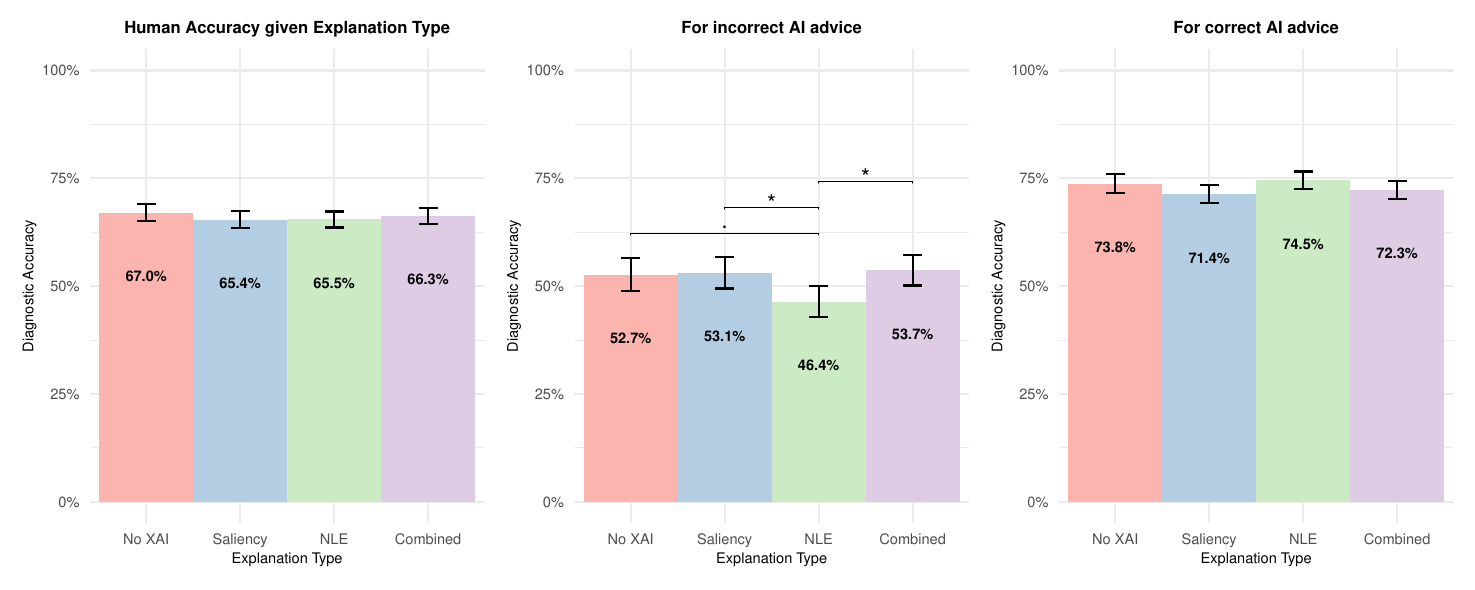}
\caption{Human accuracy given explanation types overall (left), for incorrect advice (middle), and for correct advice (right).}
\label{fig:hyp123} 
\end{figure*}

\section{Additional Main Results} \label{app:add}

In Figure~\ref{fig:hyp123}, we show the effect of explanation types (given correct and incorrect advice) on human accuracy, using our explanation classification framework. We see clearly that pure textual explanations perform much worse for incorrect advice than visual explanations.

\begin{figure*}
\centering
\includegraphics[width=1\textwidth]{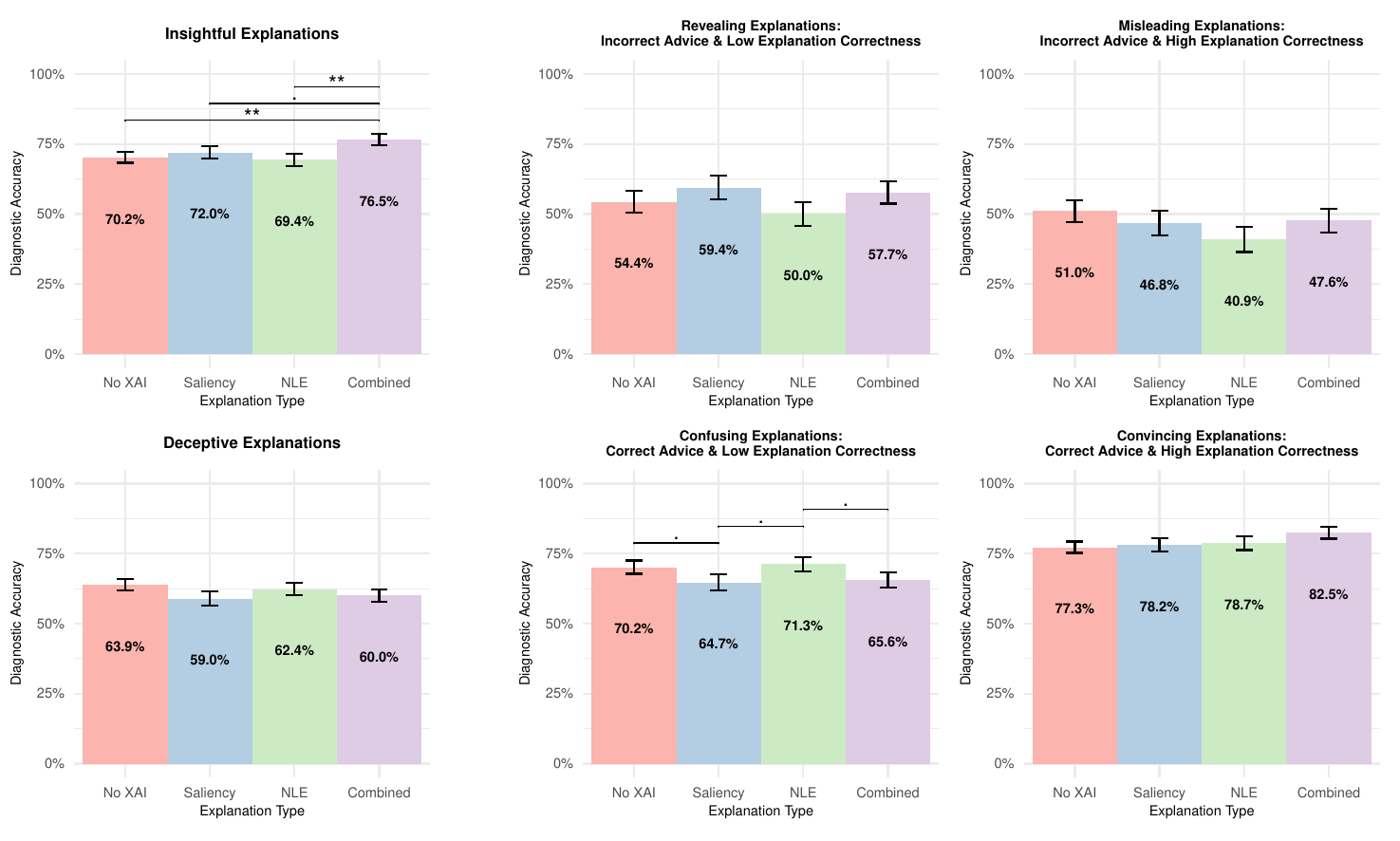} 
\caption{Multiple testing adjusted results. The bar charts and error bars represent model-based predictions of human accuracy under different conditions. $p$-values are derived from hypothesis testing, comparing human accuracy between explanation types for specific data subsets and using Benjamini-Hochberg’s corrections for multiple testing. The error bars represent standard errors. $\cdot$, *, ** ($p<0.1$, $0.05$, $0.01$)}
\label{fig:main_results_adj} 
\end{figure*}

We also include Benjamini-Hochberg's corrections for multiple testing (Figure~\ref{fig:main_results_adj}). While some effects are no longer significant, we observe that combined explanations still provide a significant boost when explanations are insightful.

    \section{Exploratory Analysis}\label{app:exploratory-analysis}

For the exploratory analysis we focus on perceived usefulness (i.e., how did participants objectively rate explanation types on a per instance level), case handling speed (how quickly they solve a case), and confidence.

\subsection{Perceived Usefulness}

Besides expressing their agreement with the AI advice, participants were also asked whether they perceived the explanation as useful. They reply via a 7-point Likert scale to the question: ``How useful was the AI model's explanation in helping you decide whether the AI was right or wrong in suggesting (e.g.) pneumonia.''. In this section we aim to understand the following: which explanation types are perceived as more useful than others, how does this interact with the correctness of the explanation, and what is the association between \emph{perceived} usefulness and the \emph{actual} usefulness, measured by the difference in their diagnostic accuracy.

In order to understand the role of perceived usefulness we consider a similar model to Equation \eqref{eq:model-definition} but instead we predict perceived usefulness $\rho_U$ and add an additional effect $A$ defined as $A=1$ when the participant agrees with the AI advice and $A=0$ otherwise.

\begin{equation}\label{eq:model-pu}
\begin{aligned}
\rho_{ij} = & \beta_0 \\
& + \beta_a C_{AI} + \beta_t \chi + \beta_p A \\
& + \beta_{t \times a} (\chi \times C_{AI}) + \beta_{p \times a} (A \times C_{AI}) \\
& + \beta_{t \times e} (\chi \times C_\chi) + \beta_{t \times p} (\chi \times A) \\
& + \beta_{t \times e \times a} (\chi \times C_\chi \times C_{AI}) \\
& + \beta_{t \times e \times p} (\chi \times C_\chi \times A) \\
& + u_{Participant} + u_{Image}
\end{aligned}
\end{equation}

This model was validated in a similar fashion as explained in Appendix~\ref{app:model-selection}. 

\begin{figure*}
    \centering
    \includegraphics[width=\linewidth]{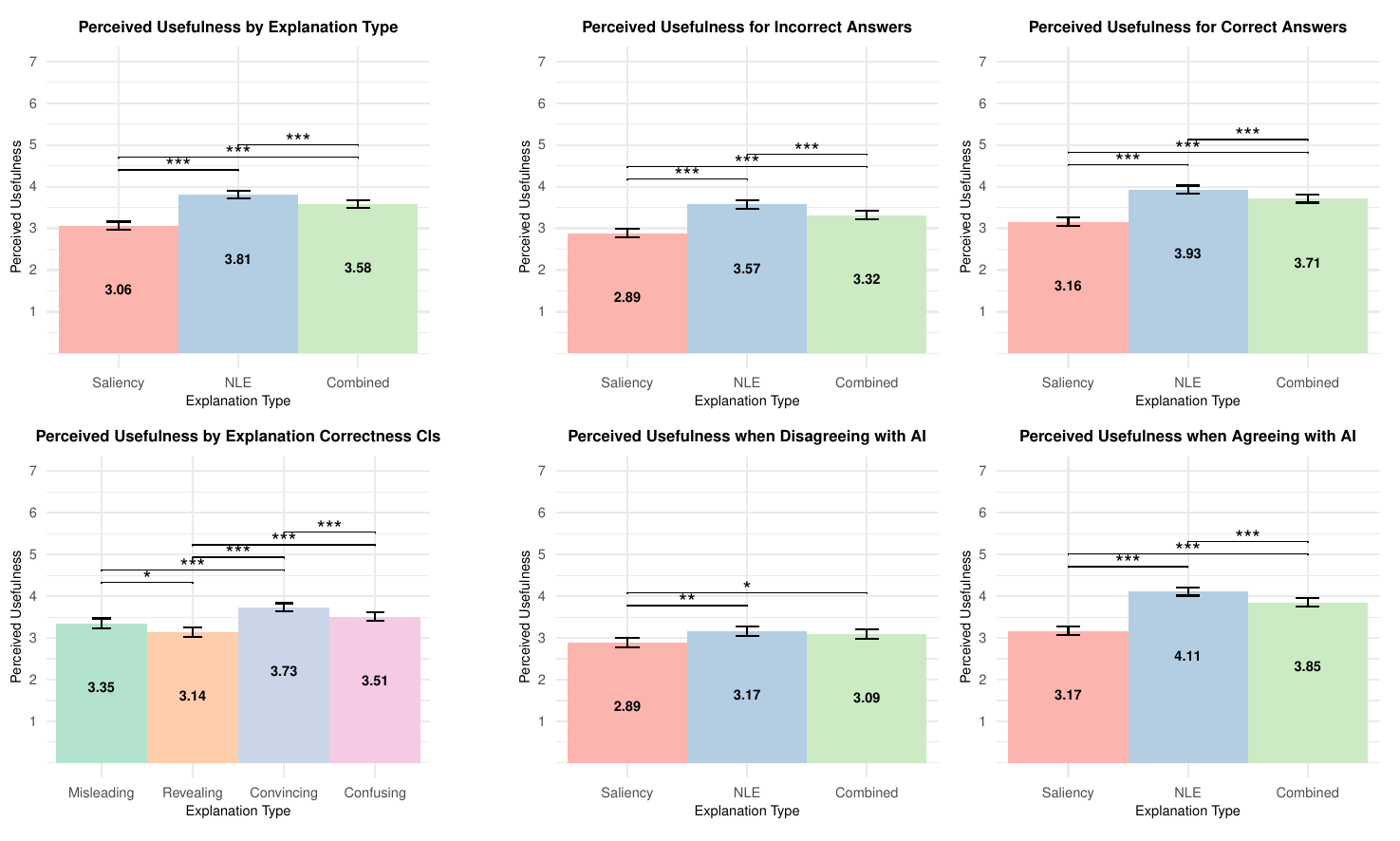} 
    \caption{Perceived usefulness $\rho_U$. The upper left shows overall $\rho_U$ with respect to explanation types. The lower left shows $\rho_U$ with respect to the explanation correctness quadrant, averaged across all types. The remaining four plots show $\rho_U$ for when participants are wrong or right, or when they agree or disagree with the AI advice. The error bars represent standard errors. $\cdot$, *, ** ($p<0.1$, $0.05$, $0.01$).}
    \label{fig:pu_main} 
\end{figure*}

\textbf{Across all scenarios NLEs obtain the highest $\rho_U$ scores}, as shown in Figure~\ref{fig:pu_main}. This is in line with our post-survey findings, where participants expressed a strong preference for NLEs. We also note that \textbf{perceived usefulness is higher both when participants agree with the AI advice (versus disagree), and when they are correct (versus when they are wrong)}. The latter suggests that perceived and actual usefulness are somewhat aligned. We also find that the difference in $\rho_U$ when disagreeing vs agreeing with the AI is significantly larger for NLEs than for saliency maps ($p<.001$), and for combined explanations than for saliency maps ($p<.001$).

In order to find out if there are significant differences between the difference in $\rho_U$ when participants are correct or wrong between explanation types, i.e., whether perceived usefulness is associated with actual usefulness more or less significantly between different explanation types, we fit a model where replace $C_{AI}$ with human accuracy in equation \eqref{eq:model-pu}. We find no significant differences between explanation types in this regard, \textbf{suggesting that perceived usefulness is equally associated with actual usefulness for all explanation types}.

\begin{figure}[ht]
    \centering
    \includegraphics[width=0.45\textwidth]{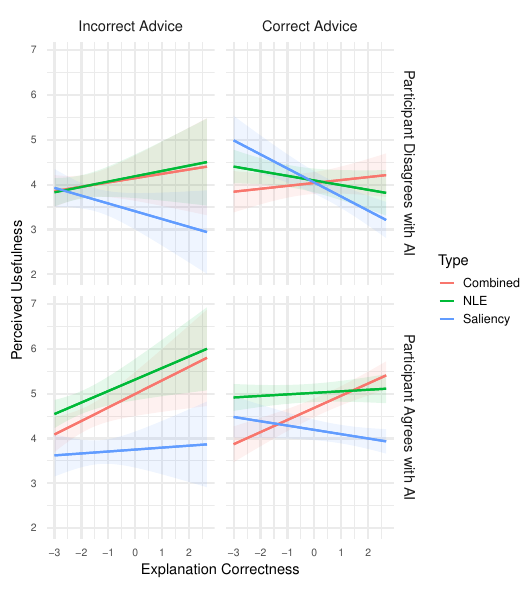}
    \caption{Perceived usefulness $\rho_U$ by AI advice correctness $C_{AI}$, user agreement $A$, and explanation correctness $C_\chi$.}
    \label{fig:pu_lines}
\end{figure}

In cases where participants correctly \emph{disagree} with the AI (top-left plot of Figure \ref{fig:pu_lines}), $\rho_U$ of saliency maps increases with decreasing explanation correctness, supporting our assumption that \textbf{incoherent saliency maps can help users detect false predictions}. This is not the case for NLEs or combined explanations. The bottom-left plot aligns with our intuition: when participants agree with the AI, even when it's wrong, they are more likely to rate factually correct explanations as useful. For the case of agreeing with correct AI advice, we observe that $\rho_U$ is by far the most highly correlated with explanation correctness for combined explanations.

\subsection{Confidence}

We consider the notion of confidence only for cases the participants rank as positive. We refer to it as \emph{positive certainty} and it's defined as the share of cases where participants rank a positive finding as ``Definitely present'', rather than ``Maybe present''. Again, we consider the same model as in \eqref{eq:model-pu}, but instead, we predict positive certainty.

\begin{equation}\label{eq:model-pc}
\begin{aligned}
\gamma_{ij} = & \beta_0 \\
& + \beta_a C_{AI} + \beta_t \chi \\
& + \beta_{t \times a} (\chi \cdot C_{AI}) + \beta_{t \times e} (\chi \cdot C_\chi) \\
& + \beta_{t \times e \times a} (\chi \cdot C_\chi \cdot C_{AI}) \\
& + u_{Participant} + u_{Image}
\end{aligned}
\end{equation}

\begin{figure*}
    \centering
    \includegraphics[width=\linewidth]{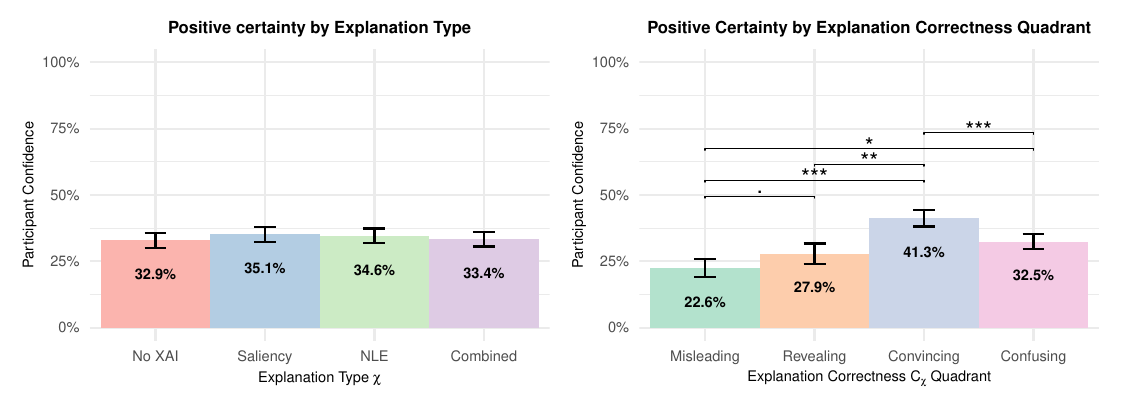} 
    \caption{Positive certainty. The left plot shows overall positive certainty with respect to explanation types $\chi$ and the right plot shows positive certainty with respect to explanation correctness $C_\chi$ quadrant. The error bars represent standard errors. $\cdot$, *, ** ($p<0.1$, $0.05$, $0.01$).}
    \label{fig:pc_charts}
\end{figure*}

This model~\ref{eq:model-pc} has a poor fit, suggesting that there is no clear relationship between positive certainty and explanation types and explanation correctness. Figure \ref{fig:pc_charts} confirms that \textbf{explanation types do not significantly affect positive certainty}. However, when subdividing into explanation correctness quadrants, we find that, unsurprisingly, \textbf{convincing explanations (correct AI advice and correct explanation) lead to the highest positive certainty}, significantly higher than all other quadrants ($p<.01$). 

\subsection{Decision Speed}

Decision speed is the time that passes between the moment participants are presented with a new case and when they enter their response. We remove cases where the time is above 2 minutes, as this likely suggests participants were interrupted (this removes 5.1\% of cases). Again, we consider the same model as in \eqref{eq:model-pu}, but we predict decision speed. We also found that adding participant agreement $A$ leads to a better fit.

\begin{equation}\label{eq:model-ds}
\begin{aligned}
\delta_{ij} = & \beta_0 \\
& + \beta_a C_{AI} + \beta_t \chi + \beta_p A \\
& + \beta_{t \times a} (\chi \times C_{AI}) + \beta_{p \times a} (A \times C_{AI}) \\
& + \beta_{t \times e} (\chi \times C_\chi) + \beta_{t \times p} (\chi \times A) \\
& + \beta_{t \times e \times a} (\chi \times C_\chi \times C_{AI}) \\
& + \beta_{t \times e \times p} (\chi \times C_\chi \times A) \\
& + u_{Participant} + u_{Image}
\end{aligned}
\end{equation}

\begin{figure*}
    \centering
    \includegraphics[width=\linewidth]{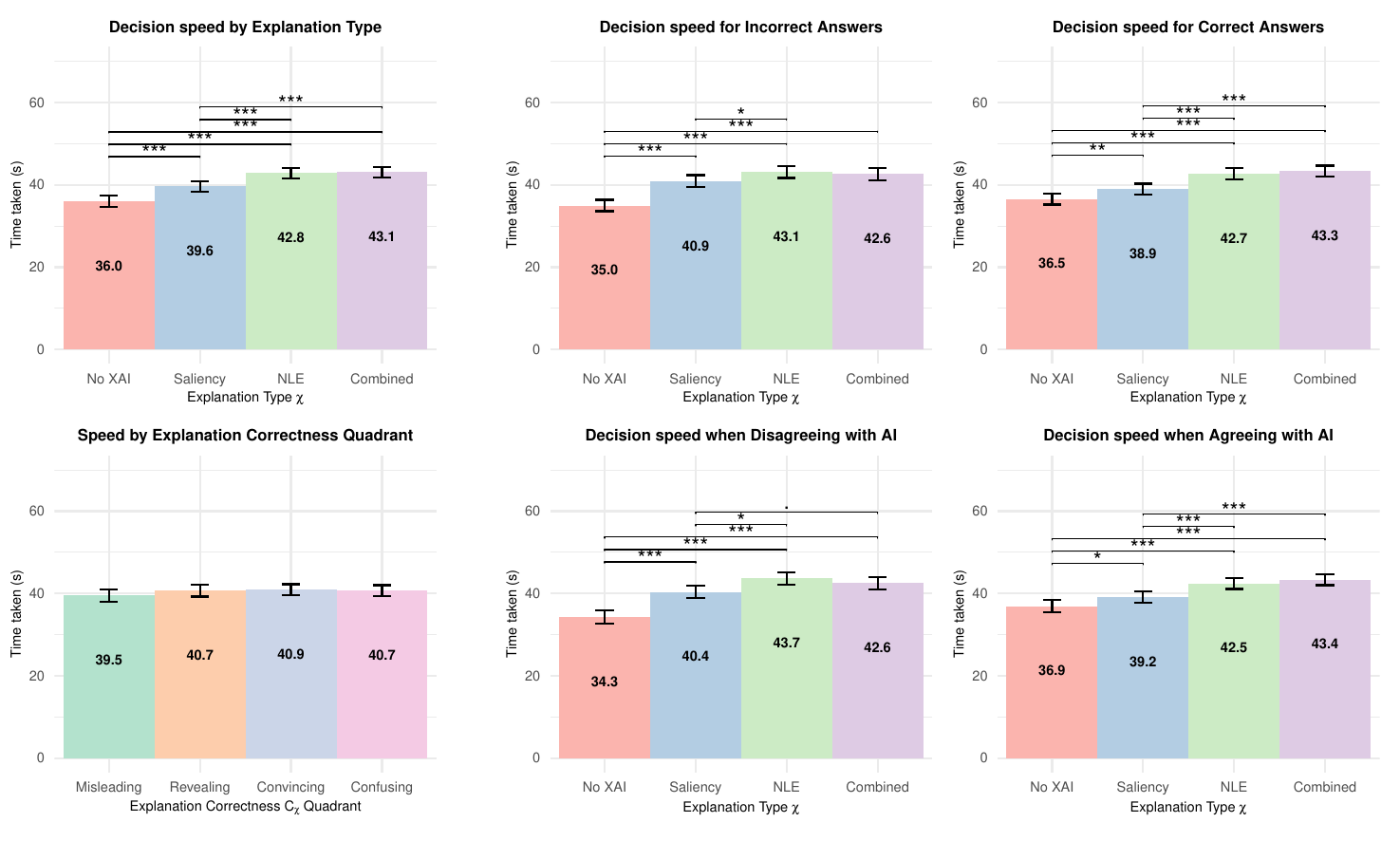}
    \caption{Decision speed. The top-left plot shows overall decision speed with respect to explanation types $\chi$ and the bottom-left plot shows decision speed with respect to explanation correctness $C_\chi$ quadrant, averaged across all types. The remaining four plots show decision speed for when participants are wrong or right, or when they agree or disagree with the AI advice. The error bars represent standard errors. $\cdot$, *, ** ($p<0.1$, $0.05$, $0.01$).}
    \label{fig:ds_charts}
\end{figure*}

The top-left plot of Figure \ref{fig:ds_charts} shows that \textbf{decision speed increases with increasing complexity of the explanation type}. There is a significant increase in time spent ($p<.001$) between each increasing complexity step, except between NLEs and combined explanations. \textbf{Time taken ranges from 36.0 seconds (no explanation) to 43.1 seconds (combined explanations)}. The duration is not significantly affected by whether participants are right or wrong, or agree or disagree with the AI. Interestingly, \textbf{explanation correctness quadrants do not show a significant effect on decision speed}. \textbf{We also find that explanation correctness has no significant effect on decision speed}, suggesting that participants do not spend more time on cases where the explanation is correct or incorrect.
    \section{Model Selection}\label{app:model-selection}

Here, we provide details on the statistical model we used to analyze our main results. The statistical model was selected based on the nature of the task and experiment design at hand and then verified using inferential statistics.

To establish the significance of our main model \eqref{eq:model-definition}, we compare it against a baseline model that disregards explanation types. The model equation is as follows:

\begin{equation}\label{eq:model-definition-baseline}
\begin{aligned}
    l_{ij} =  & \beta_0 \\
    & + \beta_a * C_{AI} \\ 
    & + u_{Participant} \\
    & + u_{Image}
\end{aligned}
\end{equation}

\paragraph{Fixed Effects.} We first select fixed effects while including random effects. As reported in the main paper, we use an LRT to test whether the added variables improve model fit. We further find the AIC (Akaike Information Criterion) is improved: from $5504.3$ to $5500.1$.

\paragraph{Random Effects.} The study design strongly suggests the inclusion of random effects $u_{Image}$ and $u_{Participant}$ as these introduce dependencies between observations. For both models, we study the random effect variances and compare the model with and without its random effects. For the baseline model \eqref{eq:model-definition-baseline} we find that $Var(u_{P})=0.056$ and $Var(u_{I})=0.400$. Further, the LRT is significant suggesting the inclusion of random effects: $\chi_{2}^2=227.86$, with $p<.0001$. We repeat this analysis for the full model \eqref{eq:model-definition}. We find $Var(u_{P})=0.059$ and $Var(u_{I})=0.295$, which are qualitatively $>0$. The LRT comparing this model with and without random effects is significant, $\chi_{2}^2=144.43$, $p<.0001$. In addition, we test incrementally only including $u_{Image}$ in comparison to a model with both random effects. Analysis of both models suggests that $u_{Participant}$ should be included. Hence, we only consider models with both random effects included.
    \section{Data Preparation} \label{app:samples}

In this section, we provide additional details on how we prepared and processed the chest X-ray cases that were included in our user study. We discuss how we obtained AI predictions, the annotation process, and then how we obtained our 80 cases from that.

\subsection{Acquiring AI Advice}

Our models perform multi-label classification, which assigns a single logit to each class. We established thresholds for each class by maximizing the Youden Index to optimize the balance between sensitivity and specificity. Upon consultations with radiologists, we selected the following subset of labels based on their clinical significance and detectability in chest X-rays alone: pneumonia, atelectasis, pulmonary edema, fluid overload/heart failure, aspiration, and alveolar hemorrhage.

\subsection{Annotation process} \label{app:annos}

The annotation process refers to the stage before running our study, where we had three expert radiologists annotate 160 examples. The radiologists classified each AI-predicted finding as \emph{Not present}, \emph{Maybe present}, or \emph{Definitely present}, based on established medical imaging standards. They also rate the correctness of NLEs and saliency maps on a 7-point Likert scale, both individually and as a combined explanation. The final values for $C_{AI}$ and $C_\chi$ for each case are obtained via majority vote and mean-centering after averaging, respectively.

When evaluating the AI advice, annotators are presented with a chest X-ray and a single class predicted by the AI (e.g. ``pneumonia''). They are then asked whether they think the class is ``Not present'' (the finding can not be seen so it is not worth mentioning or it can be mentioned negatively. For example: ``No signs of pneumonia.''), ``Maybe present'' (while the evidence is inconclusive and/or there is some ambiguity, it is worth mentioning in the radiology report that the finding may be present. For example: ``Bibasilar opacities may represent atelectasis or pneumonia.''), or ``Definitely present'' (the finding is clearly present and will be noted in the radiology report. For example: ``There are clear signs for pneumonia.''), following a common convention in evaluating the presence of chest X-ray findings (cite MIMIC-CXR, Chexpert). Both the annotators and study participants are instructed to interpret the labels as follows: 

\begin{itemize} 
    \item ``Not present'': The finding can not be seen and is therefore not worth mentioning in the radiology report (or it can be mentioned negatively). For example: ``No signs of pneumonia.''
    \item ``Maybe present'': While the evidence is inconclusive and/or there is some ambiguity, it’s worth mentioning in the radiology report that the finding may be present. For example: ``Bibasilar opacities may represent atelectasis or pneumonia.''
    \item ``Definitely present'': The finding is clearly present and will be noted in the radiology report. For example: ``There are clear signs for pneumonia.''
\end{itemize}

\begin{figure*}
\centering
\includegraphics[width=1\textwidth]{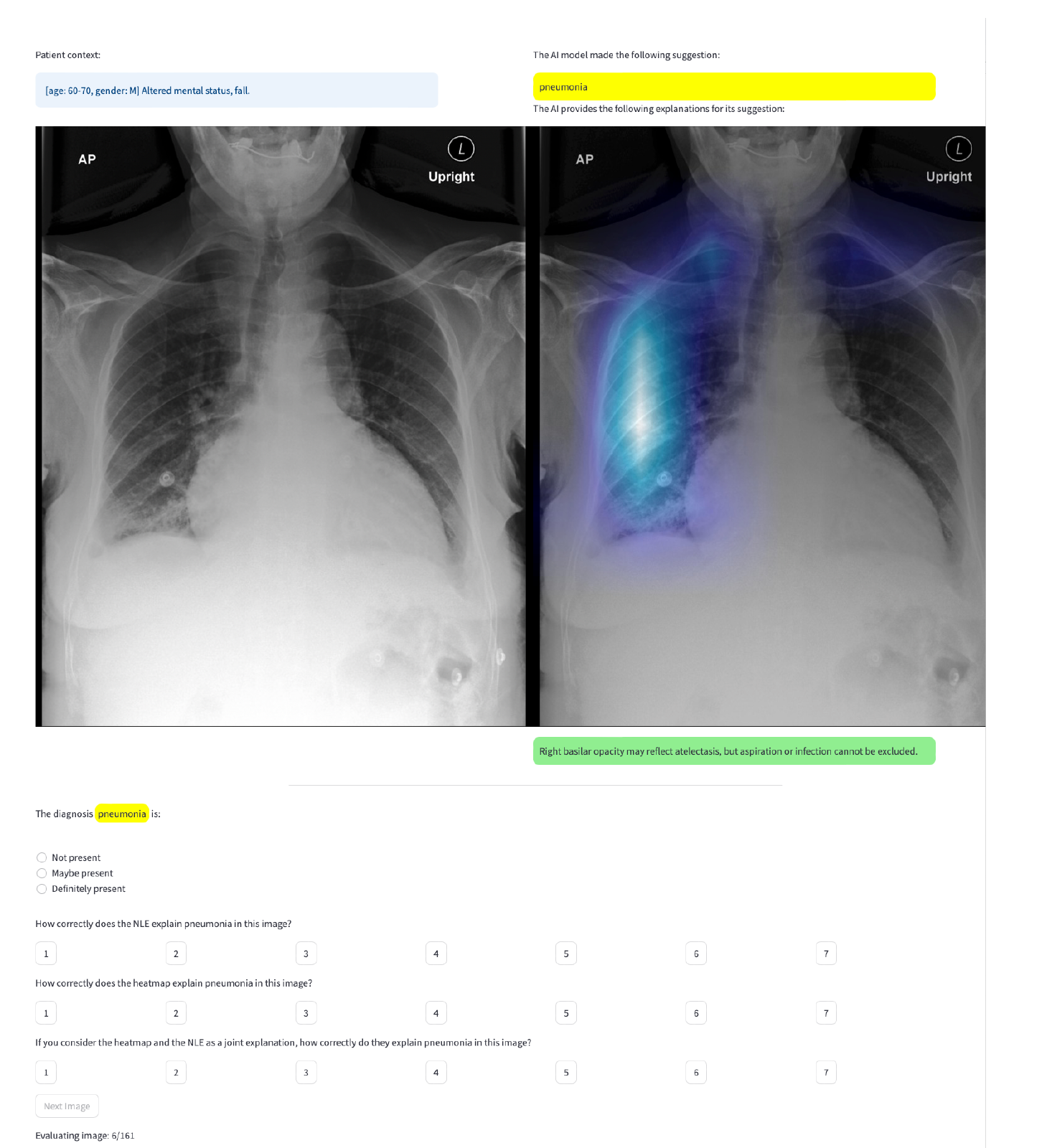} 
\caption{The user interface used by our radiologists to annotate chest x-rays.}
\label{fig:annotators} 
\end{figure*}

\begin{figure*}
\centering
\includegraphics[width=1\textwidth]{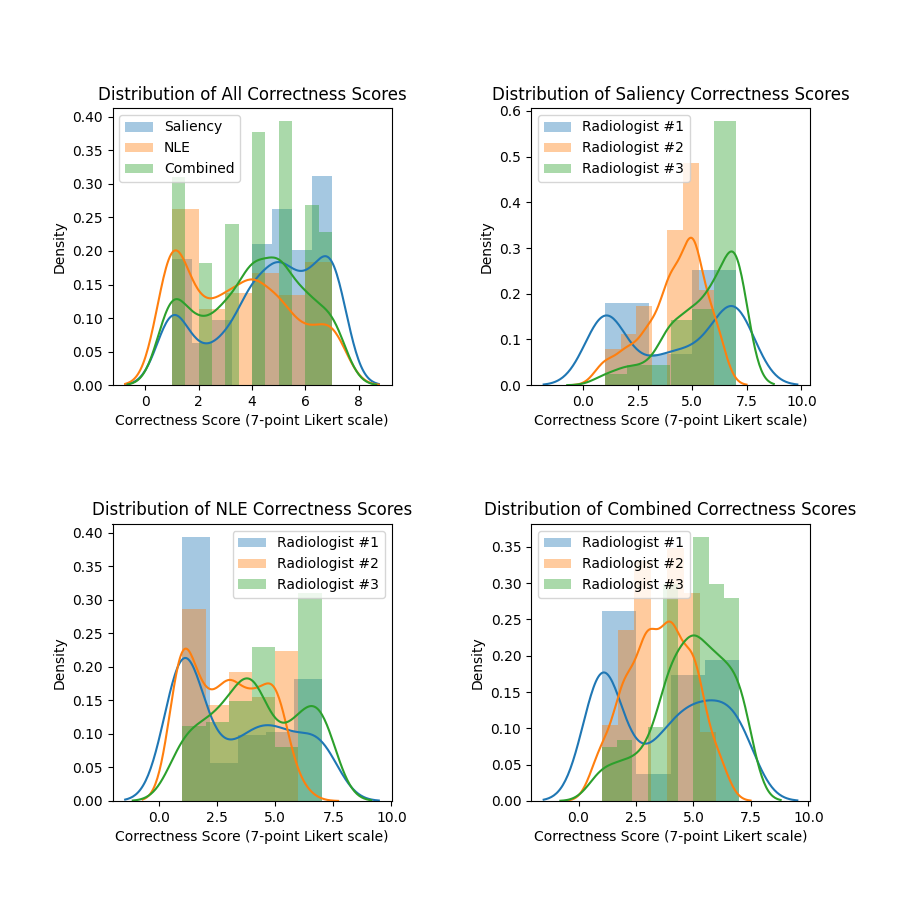}
\caption{The graphs show the distribution of explanation correctness scores $C_\chi$ assigned to the different explanation types by our annotators.}
\label{fig:ec_annos} 
\end{figure*}

The annotators also evaluate the textual explanation and saliency map for each prediction. Given that explanations can vary significantly in information richness \citet{rivera2022continuous}, we argue that a continuous scale is better suited than a binary correctness label, as has been done by \citet{morrison2023impact}. Suppose our annotators deem the AI advice (e.g. ``pneumonia'') to be correct (``Definitely present'' or ``Maybe present''). In that case, we ask them ``How correctly does the NLE (or heatmap) explain the AI advice pneumonia in this image?'' and record their response on a 7-point Likert scale. We also asked them ``If you consider the heatmap and the NLE as a joint explanation, how correctly do they explain the AI advice pneumonia in this image?'' to obtain a correctness score for the combined explanation. In case they think the AI prediction is incorrect, we still want to get a measure of how much correct information an explanation contains and ask them the following: ``How correctly does the heatmap (or NLE) highlight radiographic findings that would be relevant for the AI advice \textit{pneumonia} in this image?''. Figure~\ref{fig:ec_annos} shows the distribution of explanation correctness scores $C_\chi$. As can be seen, saliency maps are generally ranked higher than NLEs. An illustration of the annotator interface can be found in Figure~\ref{fig:annotators}.

We obtain our consensus by selecting the overall \emph{advice correctness} $C_{AI}$ as the majority vote of the three annotations, and the \emph{explanation correctness} $C_\chi$ score of each explanation as the average of the three scores. We mean-center $C_\chi$ for each type of explanation to facilitate our statistical modeling.

\subsection{Selecting 80 cases}

\begin{figure*}
\centering
\includegraphics[width=1\textwidth]{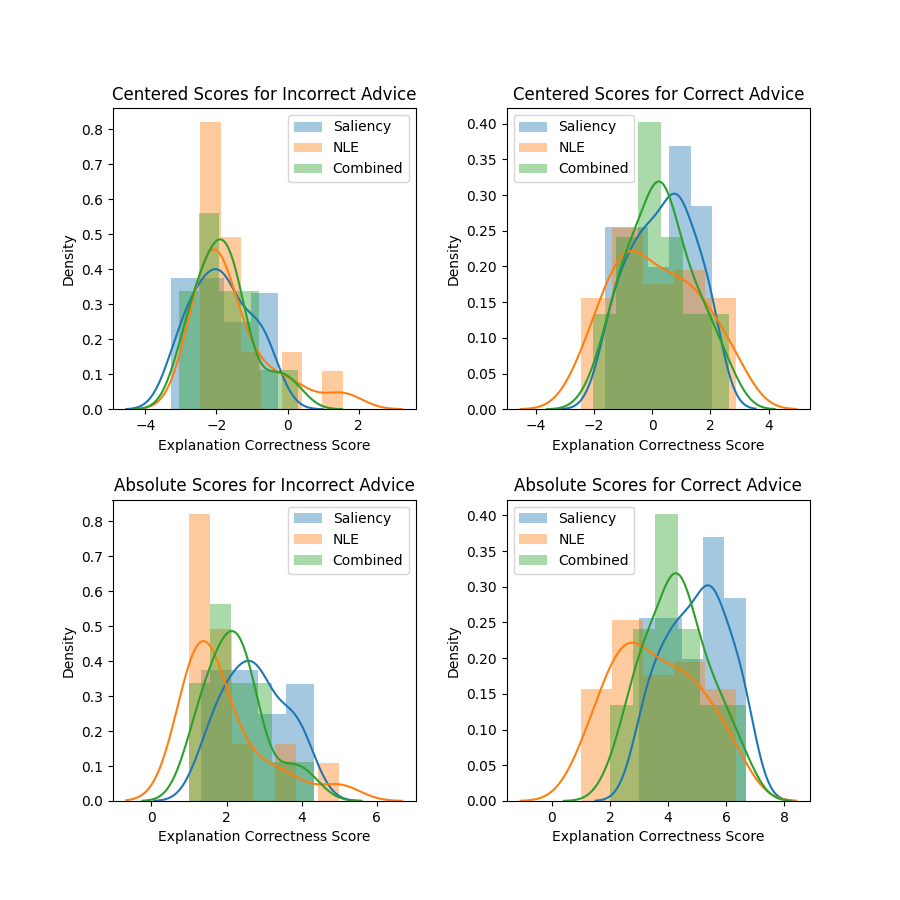}
\caption{An illustration of the distribution of explanation correctness scores included in the study.}
\label{fig:ec_scores} 
\end{figure*}

We annotated 160 cases, from which we carefully selected 80 cases that have a similar distribution of correct and incorrect AI predictions across all our classes. We also excluded ambiguous cases with significant annotator disagreement, i.e., when a case was annotated with both ``Not present'' and ``Definitely present''. Additionally, we sample examples such that the distribution of $C_\chi$ scores is as uniform as possible. The final distributions, including mean-centering, are shown in Figure~\ref{fig:ec_scores}. As expected, $C_\chi$ for positive predictions is much higher than for negative predictions.

For our selected sample we obtain pairwise kappa scores of $0.451$, $0.458$, and $0.502$ between the three annotators when grouping `Maybe present'' and ``Definitely present'' as positive (i.e., ``moderate'' agreement). Note that if we leave out ``Maybe present'' votes, we get perfect kappa scores because of the ambiguity exclusion criteria.

\begin{figure*}
\centering
\includegraphics[width=1\textwidth]{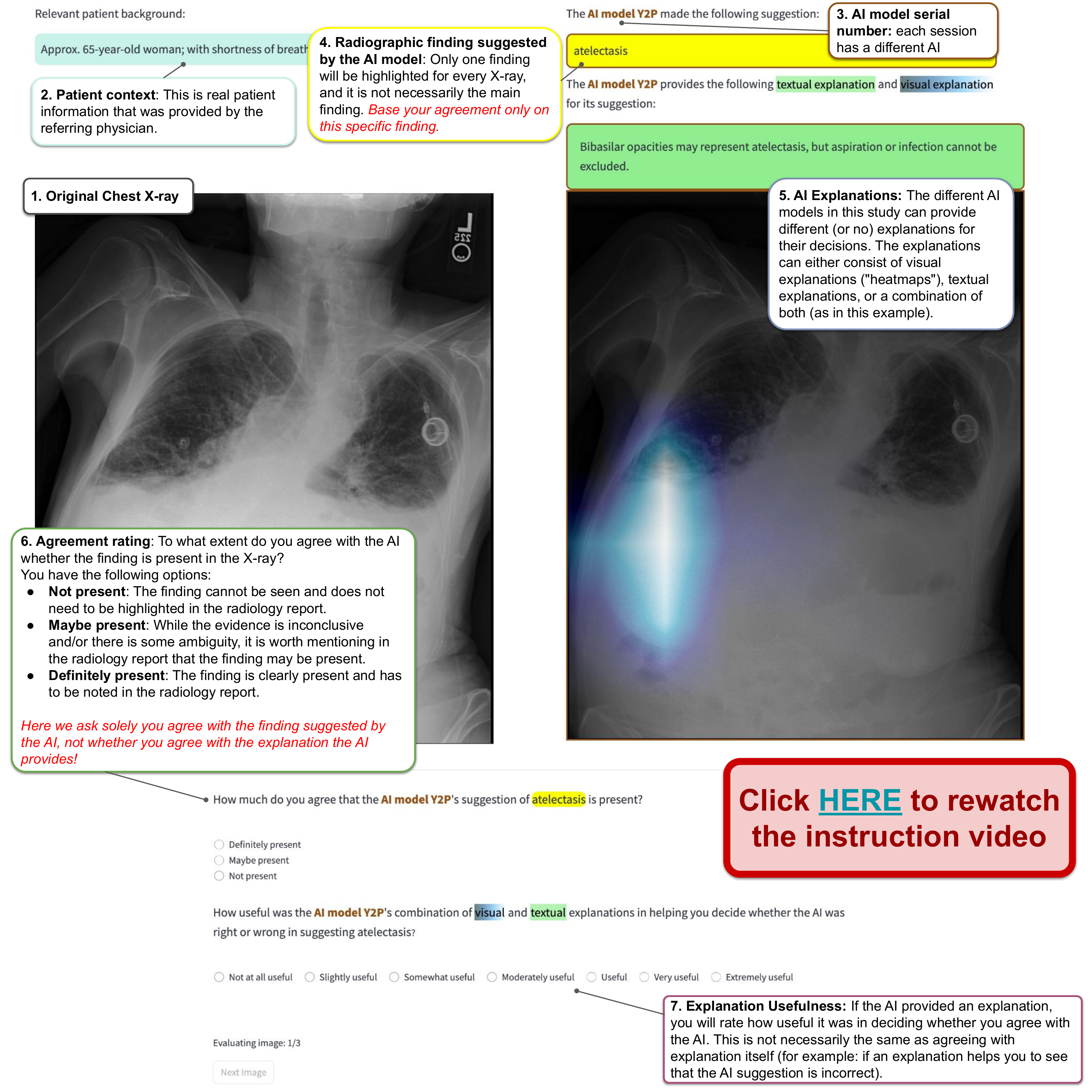}
\caption{The instruction PDF that people have access to throughout the study, which overlays instructions onto the actual study UI.}
\label{fig:screen2} 
\end{figure*}

\subsection{Distributing cases across participants and tasks} 

These 80 images were evenly distributed across four tasks and multiple participants, ensuring each image was equally represented across all tasks. This method prevents task-specific biases and maintains a consistent 70\% accuracy rate for AI advice across different explanation types.

    \section{Selected Participants} \label{app:participants}

\begin{figure*}
\centering
\includegraphics[width=1\textwidth]{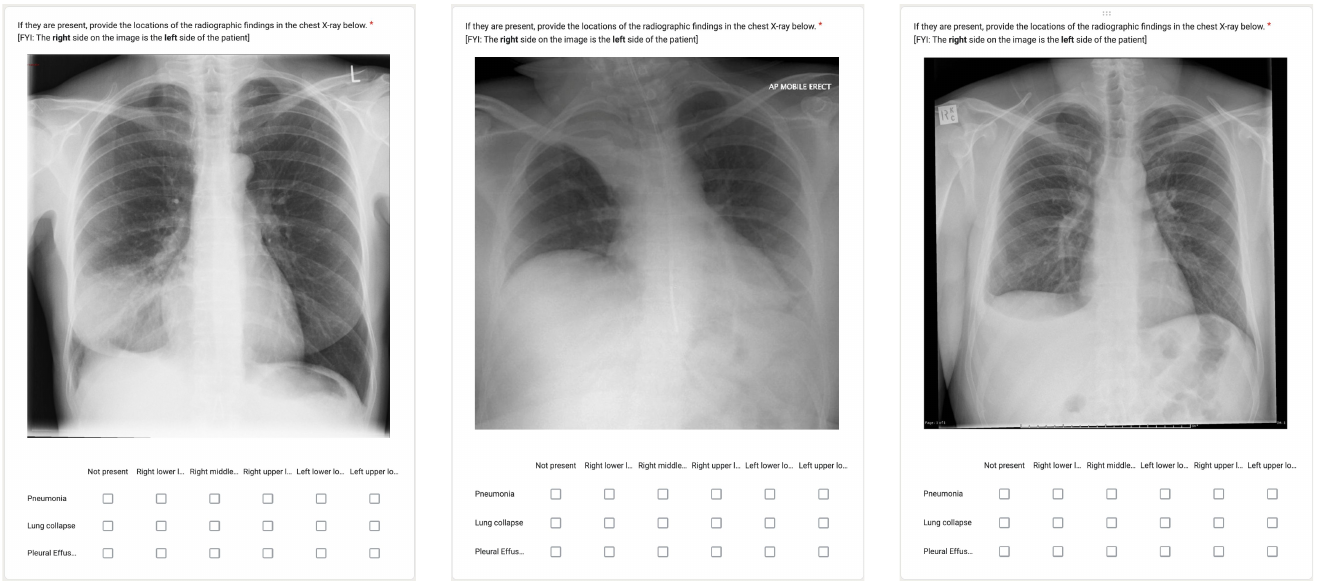}
\caption{The three test cases included in the screening survey. These X-rays contain examples of pneumonia, pleural effusion, and lobe collapse, which are the most common classes in the dataset.}
\label{fig:screen1} 
\end{figure*}

Our primary target group for this study are medical students and doctors who have undergone training in reading chest X-rays, but who are not specialist radiologists. This includes radiologists in training. We validate participants' radiology proficiency via a screening form, which contains a self-assessment as well a quiz on three chest X-rays that fulfil the medical student curriculum of the Royal College of Radiologists (UK) (an example is shown in Figure~\ref{fig:screen1}). To determine the sample size, we ran four pilot studies and used the estimated effects to run a power analysis using the model described in equation~\ref{eq:model-definition}. We found that 80 participants should provide significant power. We ended up recruiting 85 participants, as we sent out extra invitations to account for dropouts. In total, 223 people filled out our form with the three evaluation cases. Our participants range from medicine students to radiology residents (see detailed characteristics in Appendix~\ref{app:participants}). We recruit participants via mailing lists and networks focusing mainly on the United Kingdom and Romania. Participants are compensated for their time with a voucher worth the equivalent of \$38 for the one-hour study. 

\begin{figure*}
\centering
\includegraphics[width=1\textwidth]{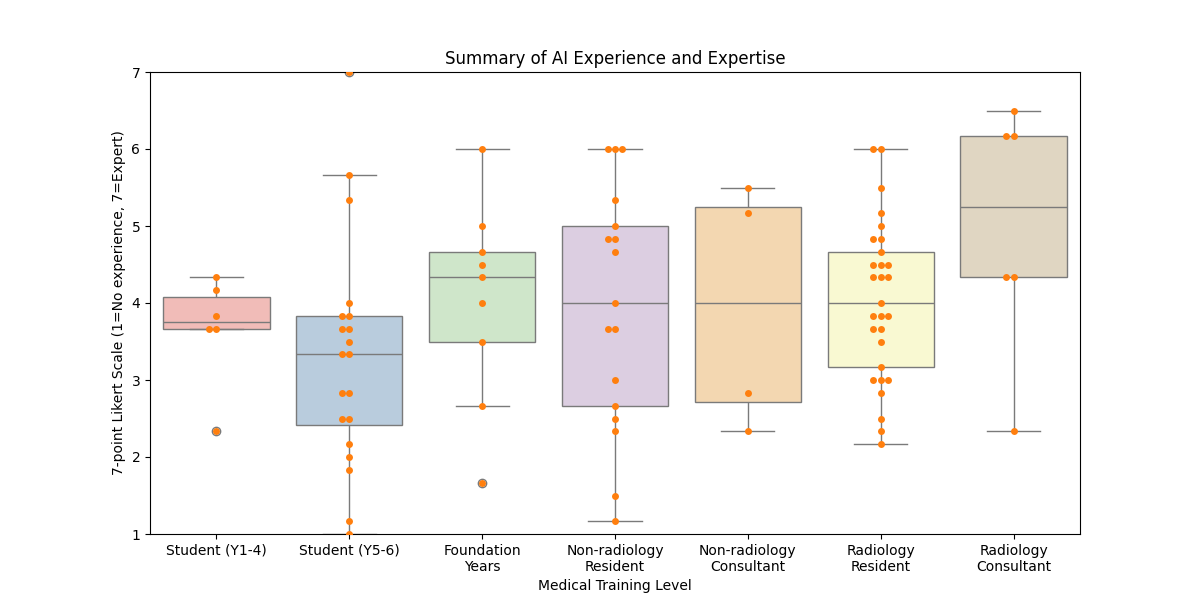}
\caption{Self-assessed levels of experience and expertise in AI (summarized across computer vision, NLP, explainable AI, and clinical decision-support systems) for different medical training levels. The questions we asked are listed in Appendix~\ref{app:ai-exp-quest}. The plot shows box plots and all individual datapoints (orange). Y$N$ is the year of medical school. Foundation years refer to the general training right after medical school (two years in the UK).}
\label{fig:ai_experience} 
\end{figure*}

\begin{figure}[ht]
\centering
\includegraphics[width=0.5\textwidth]{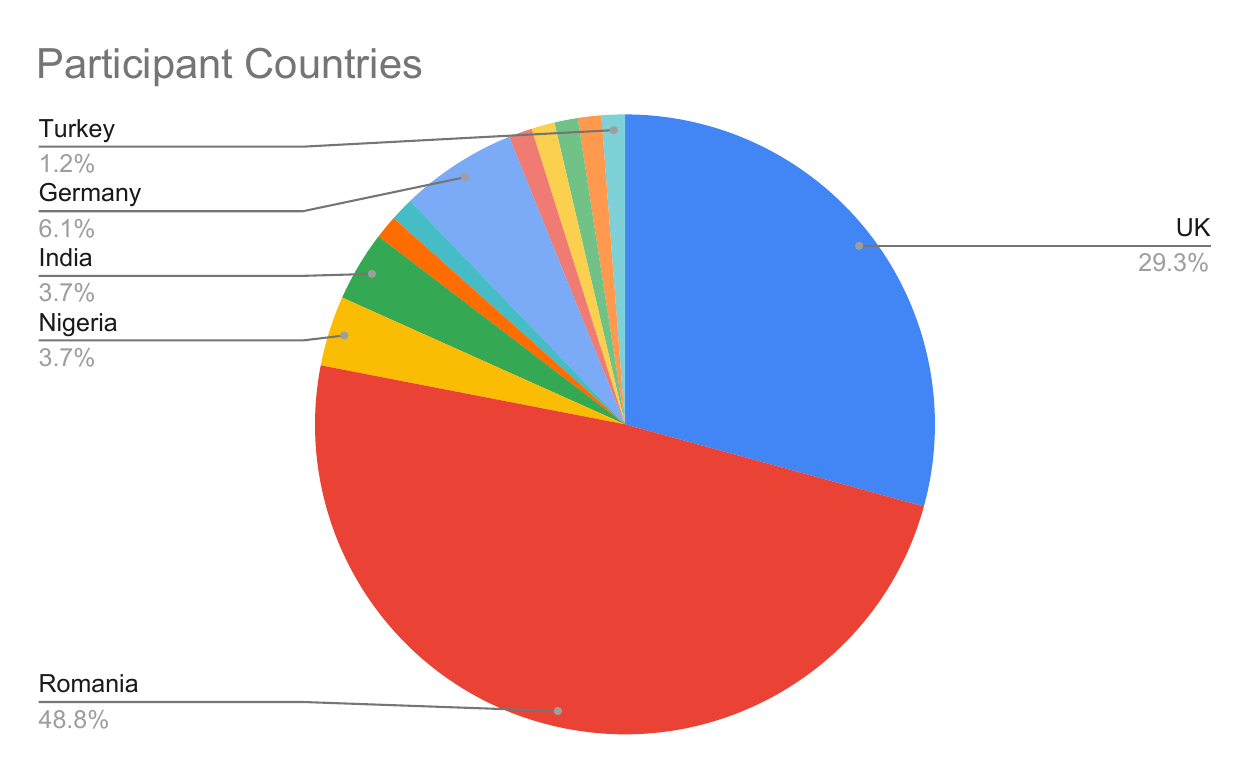}
\caption{Countries where participants have spent the most time ``studying or practising'' medicine.}
\label{fig:origin}
\end{figure}

\begin{figure}[ht] 
\centering
\includegraphics[width=0.5\textwidth]{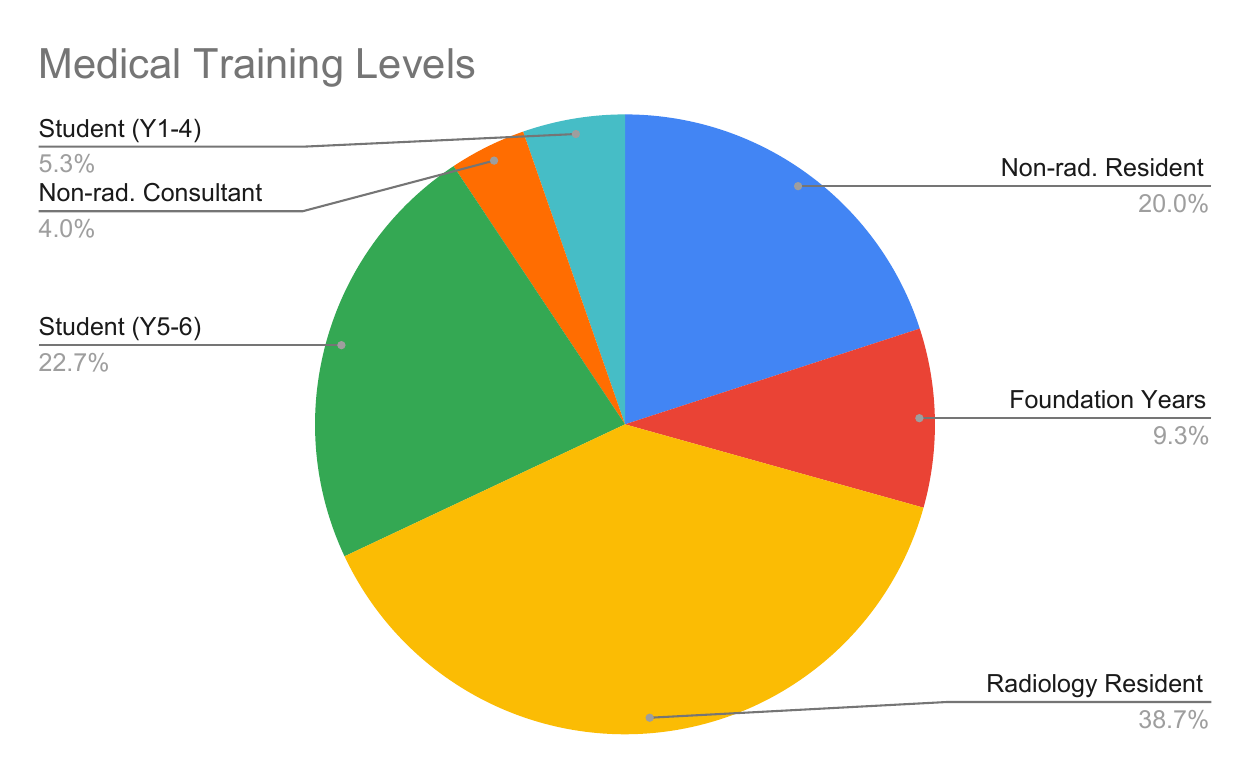}
\caption{Medical Training Level of Participants. Y$N$ is the year of medical school. Foundation years refer to the general training right after medical school (two years in the UK).}
\label{fig:pie_training}
\end{figure}

We provide descriptive information on the 85 participants included in this study. Figure~\ref{fig:screen1} shows the three test cases that we used to filter out participants for this study. Figure~\ref{fig:ai_experience} shows that self-assessed familiarity with AI technologies slightly increases with medical seniority. Very few participants rank themselves very low on this. Figure~\ref{fig:origin} gives an overview of the geographic distribution of our participants. Most participants are from UK and Romania. While developed nations are over-represented, there is a degree of diversity in the development status of the included nations. Figure~\ref{fig:pie_training} shows the distribution of medical training levels. 

    \section{Data and Subject Exploration}

\begin{figure}[ht]
\centering
\includegraphics[width=0.5\textwidth]{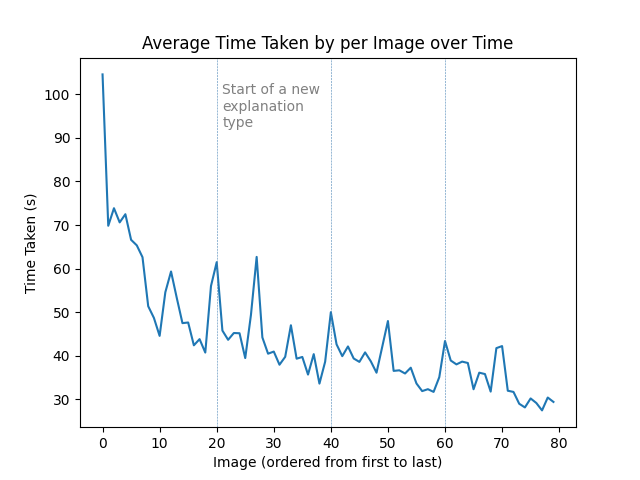}
\caption{This plot shows the average decision speed (time taken per image) and how it changed over time. The overall trend is that participants become faster over time. We can also see spikes at the start of each new task, when they are introduced to a new explanation type.}
\label{fig:tot}
\end{figure}

\begin{figure*}
\centering
\includegraphics[width=1\textwidth]{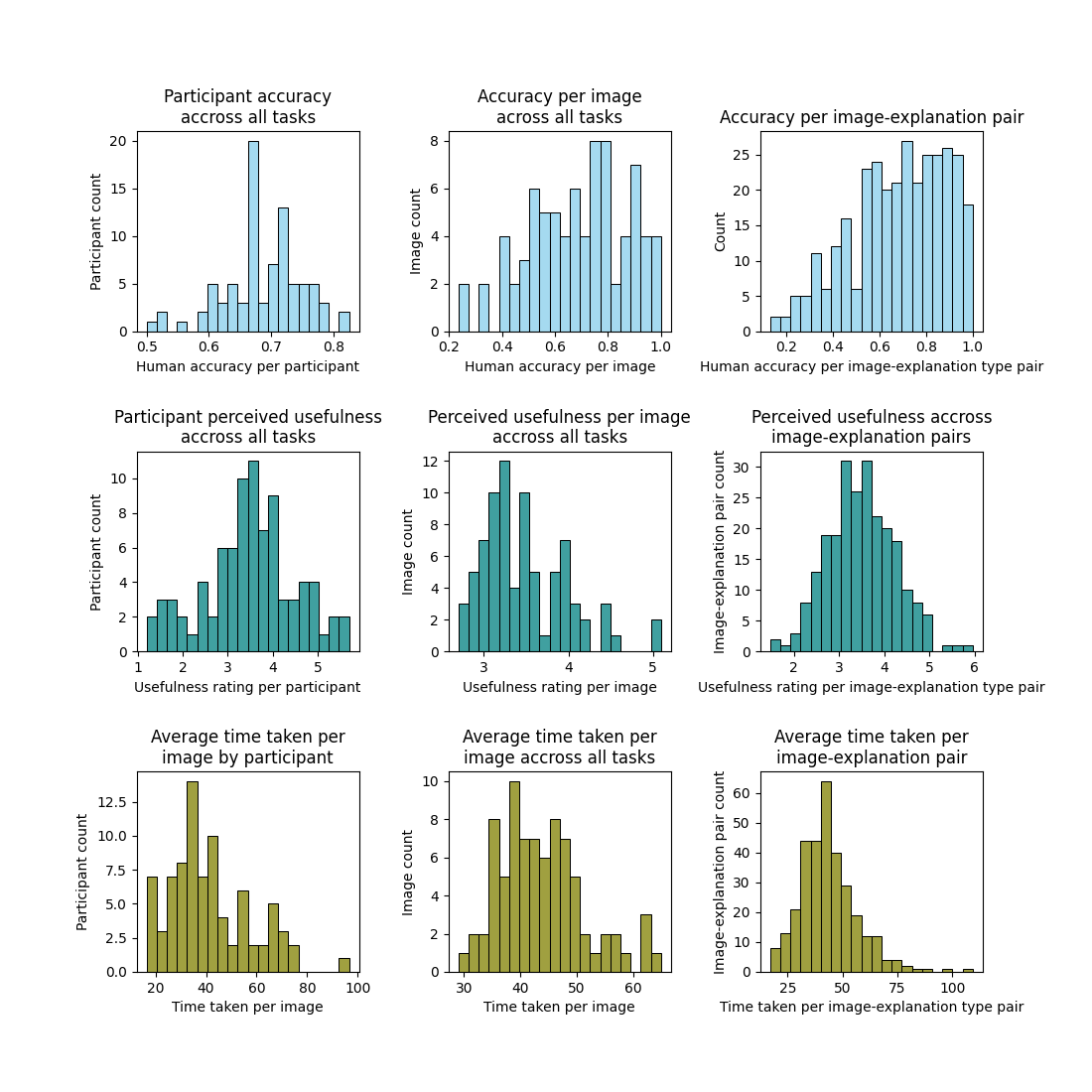}
\caption{This 3x3 plot illustrates the distributions of accuracies, perceived usefulness, and decision speed by: participant, image, and image-explanation pairing.}
\label{fig:insights} 
\end{figure*}

\begin{figure}[ht]
\centering
\includegraphics[width=0.5\textwidth]{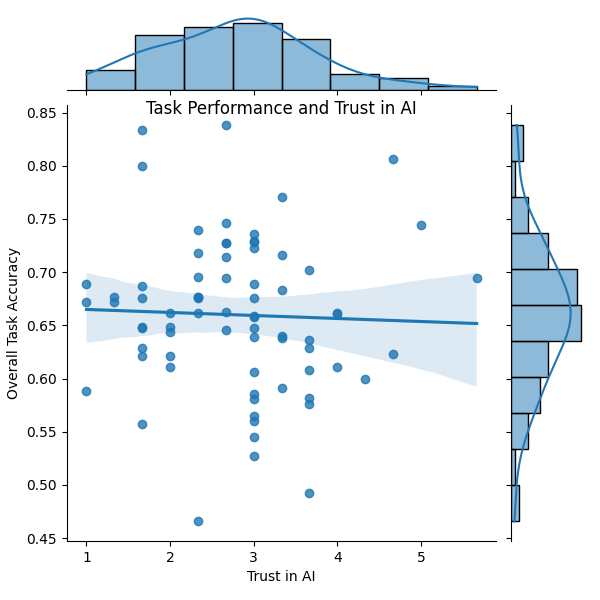}
\caption{A participant's trust in AI compared to their diagnostic accuracy across all tasks.}
\label{fig:tvp} 
\end{figure}

This section (Figures~\ref{fig:tot} to ~\ref{fig:tvp}) contains further insights into how subjects behaved during our study.

\begin{figure}[ht]
\centering
\includegraphics[width=0.5\textwidth]{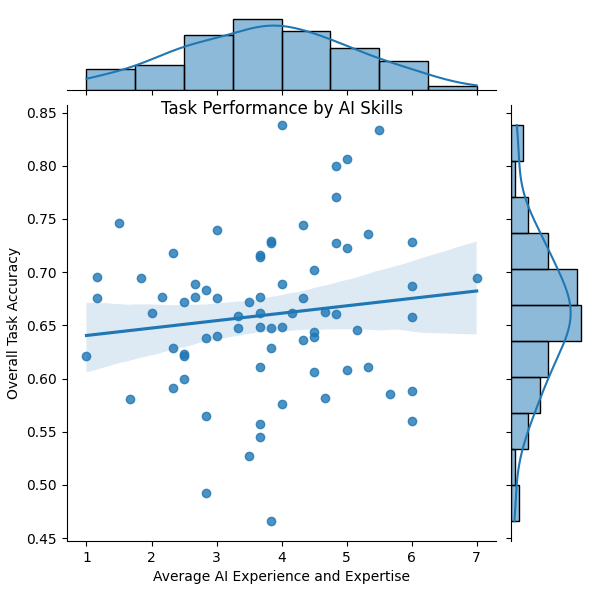}
\caption{A participant's AI experience and understanding compared to their diagnostic accuracy across all tasks.}
\end{figure}
\begin{figure}[ht]
\centering
\includegraphics[width=0.5\textwidth]{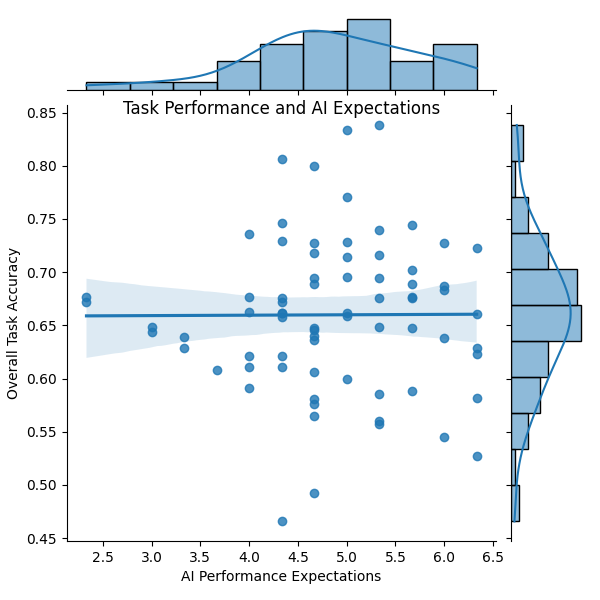}
\caption{A participant's expectation of AI compared to their diagnostic accuracy across all tasks.}
\end{figure}
\begin{figure}[ht]
\centering
\includegraphics[width=0.5\textwidth]{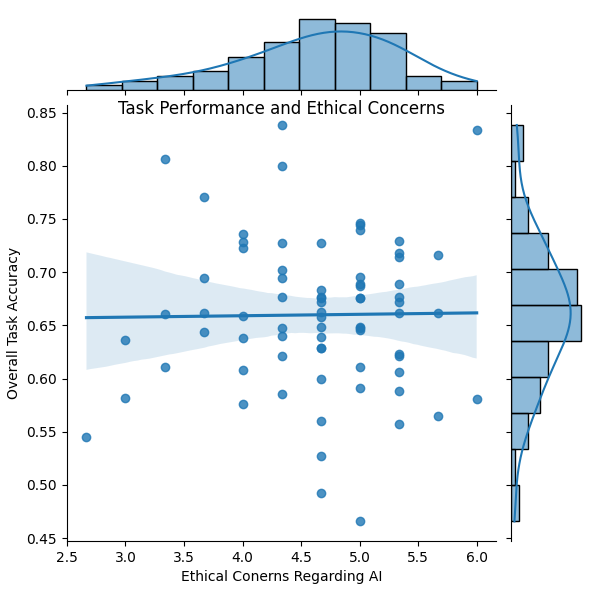}
\caption{Participant's level of ethical concerns regarding AI compared to their diagnostic accuracy across all tasks.}
\end{figure}

    \section{Participant Survey} \label{app:survey}

\subsection{Questions about level of AI expertise} \label{app:ai-exp-quest}

Participants have to agree to each of the following statements on a 7-point Likert scale from ``Strongly Disagree'' to ``Strongly Agree''.

\begin{itemize}
    \item I understand the principles behind computer vision models (i.e., AI algorithms used for analysing images) and how they work.
    \item I am familiar with language models (i.e. AI algorithms used to understand and generate language) and how they work.
    \item I understand the concepts of explainable AI (XAI), i.e., methods that try to make AI algorithms' decision-making more transparent (for example: heatmaps).
    \item I regularly use AI-powered chat tools (e.g. ChatGPT).
    \item I regularly interact with methods that make AI algorithms' decision-making more transparent.
    \item I regularly use AI-based decision-support tools for medical imaging.
\end{itemize}

\subsection{Questions about attitude towards AI}

Below are the 9 statements that were used to evaluate participants' attitudes towards AI in terms of trust, ethical concern, and performance expectations. We use the same Likert scale as above.

\paragraph{Trust}
\begin{itemize}
    \item I'm not comfortable using an AI if I don’t fully understand how it makes a decision.
    \item The use of AI should always be accompanied by the option for human review and intervention.
    \item I trust AI-based recommendations as much as those from human experts in a clinical setting.
\end{itemize}

\paragraph{Ethical Concerns}
\begin{itemize}
    \item I am not concerned about the ethical implications of using AI in healthcare.
    \item Due to the dangers of AI, its adoption should be minimised.
    \item The development of AI in healthcare should be tightly regulated.
\end{itemize}

\paragraph{Performance Expectations}
\begin{itemize}
    \item It won’t take long until AI will drastically transform healthcare.
    \item AI in its current form is still far from being ready to be used in clinical practice.
    \item I believe AI can improve the accuracy of diagnoses in healthcare.
\end{itemize}

\subsection{Explanation Type Feedback Questionnaire}

To capture participants' objective feedback on explanation types we asked the following questions for each type (only the ``trust'' question for ``No XAI'').

\begin{itemize}
    \item I trusted this AI.
    \item The explanations that were provided for the diagnoses were difficult to understand.
    \item It was transparent to me how the AI came to a diagnosis.
    \item I didn't rely on the AI’s explanations to decide whether I agree with the diagnosis or not.
    \item I have learned something from the AI’s explanations and they helped me become more proficient in reading chest X-rays.
    \item How accurate do you think this AI was (in \%)?
\end{itemize}

For all but the last question users had to respond on the same 7-point Likert scale as described above.
    
\end{document}